\documentclass[aps,prd,onecolumn,preprintnumbers,floatfix,showpacs, nofootinbib,superscriptaddress,notitlepage]{revtex4-1}

\usepackage{amsmath}           		 
\usepackage{amssymb}           		 
\usepackage{color}             		 
\usepackage{bbm}               		 
\usepackage{braket}            		 
\usepackage{xspace}            		 
\usepackage{mciteplus}         		 
\usepackage{mathtools}         		 
\usepackage{graphicx}          		 
\usepackage{slashed}           		 
\usepackage{enumerate}          		 
\usepackage[export]{adjustbox} 		 
\usepackage[english]{babel}    		 
\usepackage[utf8]{inputenc}    	 	 
\usepackage{pgffor}            		 
\usepackage{verbatim}        		 
\usepackage{tabularx}         		 
\usepackage[caption = false]{subfig} 
\usepackage{hyperref}          		 

\renewcommand{\a}{\mathbf{a}}
\renewcommand{\b}{\mathbf{b}}

\renewcommand{\k}{\mathbf{k}}
\newcommand{\p}{\mathbf{p}}
\newcommand{\q}{\mathbf{q}}

\renewcommand{\P}{\mathbf{P}}

\newcommand{\0}{\mathbf{0}}

\newcommand{\2}{\mathbf{2}}
\newcommand{\3}{\mathbf{3}}

\newcommand{\n}{\mathbf{n}}
\newcommand{\m}{\mathbf{m}}

\newcommand{\Zbb}{\mathbbm{Z}}

\newcommand{\Bc}{\mathcal{B}}
\newcommand{\Cc}{\mathcal{C}}
\newcommand{\Dc}{\mathcal{D}}

\newcommand{\Gc}{\mathcal{G}}

\newcommand{\Ic}{\mathcal{I}}

\newcommand{\Kc}{\mathcal{K}}
\newcommand{\Lc}{\mathcal{L}}
\newcommand{\Mc}{\mathcal{M}}

\newcommand{\Oc}{\mathcal{O}}
\newcommand{\Pc}{\mathcal{P}}

\newcommand{\Rc}{\mathcal{R}}

\newcommand{\Tc}{\mathcal{T}}

\newcommand{\Xc}{\mathcal{X}}

\DeclareMathOperator{\re}{Re}
\DeclareMathOperator{\im}{Im}

\newcommand{\wt}[1]{\widetilde{ #1 }}

\newcommand{\bh}[1]{\mathbf{\hat{ #1 }}}

\newcommand{\cf}{cf.\xspace}
\newcommand{\eg}{e.g.\xspace}

\newcommand{\ie}{i.e.\xspace}

\newcommand{\ansatz}{\textit{ansatz}\xspace}

\newcommand{\df}{\textrm{df}}
\newcommand{\nn}{\nonumber}
\newcommand{\diff}{\textrm{d}}

\newcommand{\pv}{\textrm{pv}}

\newcommand{\beq}{\begin{equation}}
\newcommand{\eeq}{\end{equation}}

\newcommand{\kev}{\ensuremath{{\mathrm{\,ke\kern -0.1em V}}}\xspace}
\newcommand{\mev}{\ensuremath{{\mathrm{\,Me\kern -0.1em V}}}\xspace}
\newcommand{\gev}{\ensuremath{{\mathrm{\,Ge\kern -0.1em V}}}\xspace}
\newcommand{\tev}{\ensuremath{{\mathrm{\,Te\kern -0.1em V}}}\xspace}

\usepackage{mfirstuc} 
\newcommand{\addReviewer}[2]{
  \expandafter\newcommand\csname #1\endcsname[1]{{\sf \color{#2} {#1}:\,##1}}
  \expandafter\newcommand\csname #1cor\endcsname[2]{{\color{#2} {#1}:\,\st{##1}{\sf ##2}}}
  \expandafter\newcommand\csname #1color\endcsname{#2}
}

\usepackage{soul,color}
\definecolor{chromeyellow}{rgb}{1.0, 0.65, 0.0}
\definecolor{DodgeBlue}{rgb}{0.118, 0.565,1.000}
\definecolor{asparagus}{rgb}{0.53, 0.66, 0.42}
\definecolor{cadmiumgreen}{rgb}{0.0, 0.42, 0.24}

\definecolor{jlab_red}{RGB}{192,39,45}
\definecolor{jlab_orange}{RGB}{249,102,0}
\definecolor{jlab_blue}{RGB}{47,122,121}
\definecolor{jlab_green}{RGB}{65,125,10}

\addReviewer{rb}{jlab_blue}
\addReviewer{aj}{jlab_orange}
\addReviewer{mh}{jlab_red}

\hypersetup{
    pdfnewwindow=true,         		 
    colorlinks=true,           		 
    linkcolor=blue,            		 
    citecolor=blue,            		 
    filecolor=blue,            		 
    urlcolor=blue              		 
}

\newcommand{\jlab}{
	Thomas Jefferson National Accelerator Facility,
	12000 Jefferson Avenue, 
	Newport News, VA 23606, USA
}
\newcommand{\odu}{
	Department of Physics,
	Old Dominion University,
	Norfolk, Virginia 23529, USA
}

\begin{document}

\title{Three-body scattering and quantization conditions \\ from $S$ matrix unitarity}


\author{Andrew W. Jackura}
\email[e-mail: ]{ajackura@jlab.org}
\affiliation{\jlab}
\affiliation{\odu}

\preprint{JLAB-THY-22-3664}

\begin{abstract}
Two methodologies have been presented in the literature which connect relativistic three-particle scattering amplitudes with lattice QCD spectra -- the ``relativistic effective field theory'' approach and the ``finite-volume unitarity'' method. While both methods have been shown to be equivalent in various works, it has not been shown how to arrive at the relativistic effective field theory results directly from $S$ matrix unitarity. In this work, we provide a simple proof of the relativistic effective field theory form of the scattering equations directly from unitarity. Motivated by the finite-volume unitarity approach, we then postulate a set of quantization conditions which relate the finite-volume energy spectra to the $K$ matrices which drive the short-distance physics in the scattering equations, obtaining all previously known results for three identical particles. This work also presents new relations which provide a new pathway to  generalize the results to arbitrary systems.
\end{abstract}
\date{\today}
\maketitle
%

\section{Introduction}
\label{sec:introduction}

Calculating few-body dynamics from first principles Quantum Chromodynamics (QCD) has become increasingly necessary to resolve many outstanding problems in nuclear and particle physics. Algorithmic and theoretical advancements in lattice QCD have made it possible to compute a number of non-perturbative observables relevant for the study of low-energy phenomena. While lattice QCD is the central tool for accessing strongly interacting few-body systems, additional concepts are needed to link the resulting finite-volume observables to infinite-volume scattering and transition amplitudes. Fortunately, a synergistic approach between lattice QCD and scattering theory has resulted in a rigorous pathway to connect the two classes of observables.

The key insight, first proposed by L\"uscher~\cite{Luscher:1985dn,Luscher:1986n2,Luscher:1990ux}, involves constructing non-perturbative mappings between finite-volume observables and scattering amplitudes. These mappings, often called \emph{quantization conditions}, have been developed for any $\2\to\2$~\footnote{We adopt the notation $\n\to\m$ which indicates a scattering process with $\n$ incoming and $\m$ outgoing stable hadrons.} scattering processes~\cite{Rummukainen:1995vs, Kim:2005gf,He:2005ey,Davoudi:2011md,Hansen:2012tf,Briceno:2012yi,Briceno:2013lba, Briceno:2014oea,Romero-Lopez:2018zyy}, and have been successfully applied to access various hadronic reactions from lattice QCD~\cite{Dudek:2010ew,Beane:2011sc,Pelissier:2012pi,Dudek:2012xn,Liu:2012zya,Beane:2013br,Wilson:2014cna,Dudek:2014qha,Orginos:2015aya,Berkowitz:2015eaa,Lang:2015hza,Wilson:2015dqa,Dudek:2016cru,Briceno:2016mjc,Moir:2016srx,Bulava:2016mks,Hu:2016shf,Alexandrou:2017mpi,Bali:2017pdv,Wagman:2017tmp,Andersen:2017una,Briceno:2017qmb,Woss:2018irj,Brett:2018jqw,ExtendedTwistedMass:2019omo,Mai:2019pqr,Woss:2019hse,Wilson:2019wfr,Cheung:2020mql,Rendon:2020rtw,Woss:2020ayi,Horz:2020zvv}.~\footnote{See Ref.~\cite{Briceno:2017max} for a recent review on the subject.} In addition to standard two-particle systems, there has been headway in determining matrix elements involving external currents~\cite{Lellouch:2000pv,Kim:2005gf,Christ:2005gi,Hansen:2012tf,Briceno:2014uqa,Briceno:2015csa,Agadjanov:2016fbd,Bernard:2012bi,Briceno:2012yi,Briceno:2015tza,Baroni:2018iau,Briceno:2019nns,Briceno:2019opb,Briceno:2020vgp,Briceno:2020xxs,Briceno:2021xlc,Sherman:2022tco,Lozano:2022kfz}, with applications for decays and resonance transition amplitudes literature~\cite{Blum:2011ng,Boyle:2012ys,Feng:2014gba,Blum:2015ywa,Bai:2015nea,Briceno:2015dca,Briceno:2016kkp,Feng:2018pdq,Alexandrou:2018jbt}.

At the forefront of this venture is analyzing three-particle systems, which has seen rapid theoretical developments within the past decade.~\footnote{Recent reviews on this emerging topic include Refs.~\cite{Hansen:2019nir,Mai:2021lwb}.} Much of the research thus far has been on constructing the quantization conditions to link the three particle spectrum to $\3\to\3$ scattering amplitudes. A complication that arises is the number of degrees-of-freedom compared to the simpler two-particle case. Because of this, the framework for computing $\3\to\3$ amplitudes is considerably more challenging, involving multiple intermediary steps such the need to solve integral equations for the on-shell amplitude. 

Two methodologies have emerged in the literature to derive the quantization conditions and the relativistic integral equations.~\footnote{In addition to these apporoaches, Refs.~\cite{Polejaeva:2012ut,Guo:2017crd,Guo:2017ism,Hammer:2017uqm,Hammer:2017kms,Meng:2017jgx,Pang:2019dfe,Muller:2021uur} have looked at the problem using generic non-relativistic effective field theories. Since our focus is on the relativistic framework, we will not discuss them further.} The first is the relativistic effective field theory (RFT) method, which generates the condition by summing on-shell projected generalized Feynman diagrams to all-orders~\cite{Hansen:2014eka,Hansen:2015zga,Hansen:2016ync,Briceno:2017tce,Briceno:2018mlh,Briceno:2018aml,Briceno:2019muc,Blanton:2019igq,Hansen:2020zhy,Blanton:2020gha}. On-shell physics is driven by singularities in the physical energy region of interest, which in turn cause the leading finite-volume effects. By isolating the volume dependence, a set of quantization conditions are derived as well as a set of integral equations in the infinite-volume limit. The original derivation focused on three-identical particles, but has since been extended to include coupled $\2\to\3$ processes~\cite{Briceno:2017tce}, as well as for distinguishable scalar particles~\cite{Blanton:2020gmf,Blanton:2021mih}, three pions with any isospin~\cite{Hansen:2020zhy}, and transition amplitudes~\cite{Hansen:2021ofl}. The latest results for non-degenerate scalars use a modified form of the RFT method which uses time ordered perturbation theory (TOPT) instead of covariant perturbation theory~\cite{Blanton:2020gha,Blanton:2020gmf,Blanton:2021mih}. Applications of this framework have appeared in the literature~\cite{Briceno:2018mlh,Blanton:2019vdk,Romero-Lopez:2019qrt,Fischer:2020jzp,Jackura:2020bsk}, including the first determination of the energy-dependent $3\pi^+\to 3\pi^+$ amplitude~\cite{Hansen:2020otl}.

The second approach taken to derive the scattering and finite-volume frameworks is the finite-volume unitarity (FVU) method. Motivated in the fact that on-shell physics is responsible for the dominant finite-volume effects, the FVU approach derives an on-shell representation for the scattering amplitude directly from $S$ matrix unitarity, which constrains all possible on-shell intermediate states~\cite{Mai:2017vot,Jackura:2018xnx,Mikhasenko:2019vhk,Dawid:2020uhn}. Unitarity has long been used to reconstruct the non-analytic structure of scattering amplitudes~\cite{Eden:1966dnq,Gribov:1186219}, with applications for three-body systems primarily in the phenomenological study of hadronic decays~\cite{Fleming:1964zz,Holman:1965mxx,Aaron:1966zz,Grisaru:1966uev,Aitchison:1966lpz,Ascoli:1975mn, Mikhasenko:2018bzm}. The resulting scattering integral equation is driven by the $K$ matrix, which is an unknown object that contains all short-distance physics which cannot go on-mass-shell. The on-shell representation is then used to suggest a finite-volume counterpart, where all continuous integrations are replaced by discrete sums over quantized momenta, giving the quantization condition as the point where the finite-volume amplitude has a pole singularity~\cite{Mai:2017bge,Doring:2018xxx,Mai:2018djl}. As with the RFT formalism, the FVU equations have been used in numerical studies of various systems~\cite{Mai:2018djl,Culver:2019vvu,Horz:2019rrn,Sadasivan:2020syi,Alexandru:2020xqf}, which includes the first calculation of the $a_1(1260)$ resonance pole position from lattice QCD~\cite{Mai:2021nul}.

Although both approaches use on-shell physics as a basis for their frameworks, the resulting scattering equations and quantization conditions appear structurally different. Since their original inceptions, it has been shown that both the integral equations~\cite{Jackura:2019bmu} and the finite-volume conditions~\cite{Blanton:2020jnm} are equivalent by fixing the $\3\to\3$ amplitude as the observable and deducing relations between the different short-distance objects that appear in each framework. While equivalent, it has yet to be shown how to derive the RFT form of the equations by using $S$ matrix unitarity alone as in the FVU approach. The absence of such a direct proof has prevented a conceptual unification of both approaches, even though both rely on on-shell physics. 

Our objective here is to provide that conceptual unification, where we show how to directly access the RFT form of the integral equations and quantization conditions by considering the consequences of $S$ matrix unitarity alone. While we follow the FVU methodology, we present a modification which allows us to immediately access the scattering equations using the known results from $\2\to\2$ scattering as the structural template. Our viewpoint is that the RFT and FVU methods should not be thought of two separate approaches, but rather adjacent procedures which could be used interchangeably. Following this procedure, we generate new relations for the scattering integral equations and quantization conditions, which may prove useful for numerical applications

In this work, we strengthen the relationship between the FVU and RFT methods by using the the FVU approach to derive the RFT result. In doing so, we present a streamlined procedure to obtain the infinite-volume scattering amplitudes and finite-volume quantization conditions, which provides an outline on how it can be generalized to accommodate systems with any number of incoming and outgoing particles. We present an overview of the main results in Sec.~\ref{sec:overview}, highlighting new forms of the integral equations and quantization conditions. In Sec.~\ref{sec:scattering}, we define the $\3\to\3$ $S$ matrix elements and kinematic constraints, and give the unitarity constraint in Sec.~\ref{sec:unitarity}. From the unitarity constraint, in Sec.~\ref{sec:scattering_equations} we propose an on-shell representation for the $\3\to\3$ scattering amplitude in terms of an integral equation. A new form for the integral equation presented here, which may prove useful for numerical applications. The infinite-volume scattering equations provide the necessary structure to derive the quantization conditions in terms of the unknown $K$ matrices. 

 We then embed the system in a finite cubic volume in Sec.~\ref{sec:QC}, and write the corresponding finite-volume amplitudes which replace all on-shell singularities by their finite-volume counterparts which give the dominant volume dependence. Quantizations conditions follow by identifying the pole structure of the finite-volume amplitudes, and agree with the ones derived using the TOPT version of the RFT method. Section~\ref{sec:alternative} provides new representation of the scattering equations and quantization conditions by performing a few simple operations on the result given in Sec.~\ref{sec:scattering_equations}. Additionally, we show how this new form arises naturally from the unitarity relations of the connection $\3\to\3$ amplitude in Sec.~\ref{sec:alternative_derivation}, and provides the key physical insight to the difference between the original FVU and RFT formalisms. This new set of equations are closely related to the original framework proposed by Ref.~\cite{Hansen:2014eka,Hansen:2015zga}, and we show how to recover them by using simple symmetrization identities in Sec.~\ref{sec:symmetric_K}. Finally we summarize our approach in Sec.~\ref{sec:conclusions}.

\section{Overview}
\label{sec:overview}

Before detailing the derivation of the results, we present a summary of the main points and introduce new forms for the $\3\to\3$ integral equations and quantization conditions. We seek elements of the $S$ matrix subject to the unitarity constraint $S^{\dag} S = 1$. The $S$ matrix can be decomposed as $S = 1+iT$, where the interacting $T$ matrix has fully disconnected contributions removed.~\footnote{Fully disconnected contributions are those pieces of $S$ matrix where all particles are causally separated such that they do not interact. For $\3\to\3$ processes, the $T$ matrix contains additional disconnected terms where two of the particles interact and the third remains causally separated~\cite{Weinberg:1995mt}, which is discussed at the end of this section.} The unitarity condition for the $T$ matrix is given by
\begin{align}
	\label{eq:Tmatrix_unitarity}
	T - T^{\dag} = iT^{\dag} T \, , 
\end{align}
which constrains the analytic structure of scattering amplitudes~\cite{Eden:1966dnq}. Matrix elements of Eq.~\eqref{eq:Tmatrix_unitarity} lead to a non-linear relation between the imaginary part of the scattering amplitude and a product of amplitudes with an infinite number of intermediate processes. If we limit the allowed total energy to the elastic scattering region, in our case energies above the three-particle threshold and below the first inelastic threshold, then higher multi-particle intermediate states are kinematically inaccessible, and the unitarity condition contains only a finite number of terms.

We exploit Eq.~\eqref{eq:Tmatrix_unitarity} to construct on-shell representations of the $\3\to\3$ scattering amplitude. As detailed in Sec.~\ref{sec:scattering}, the resulting constraints for the $T$ matrix elements are satisfied by linear integral equations driven by an unknown dynamical function called the $K$ matrix,
\begin{align}
	\label{eq:Tuu_onshell}
	\Tc^{(u,u)}(\p,\k) & = \Kc_{3}^{(u,u)}(\p,\k) - \int_{p'} \int_{k'} \Kc_{3}^{(u,u)}(\p,\p') \,  \Gamma(\p',\k')  \, \Tc^{(u,u)}(\k',\k) \, ,
\end{align}
where we have written the $T$ matrix element in the $k\ell m_{\ell}$ basis, meaning two of the particles are coupled into a pair with definite angular momentum $\ell$ and projection $m_{\ell}$, while the third remains a spectator with momentum $\k$.~\footnote{The amplitude is also a function of the total momentum of the system $P$, which we leave implicit.} Both the initial and final states for all objects in Eq.~\eqref{eq:M3_onshell} are expressed in this basis, and thus the relation is a matrix equation in angular momentum space (with indices implicit), while it is an integral equation in the spectator momenta where we have introduced the shorthand notation for integrations
\begin{align}
	\int_{k} \equiv \int\!\frac{\diff^3\k}{(2\pi)^{3} \, 2\omega_k} \, . \nn
\end{align}
Explicitly, the $T$ matrix elements are defined as
\begin{align}
	\bra{P',\p,\ell' m_{\ell'}} T \ket{P,\k,\ell m_{\ell}} \equiv (2\pi)^{4}\delta^{(4)}(P' - P) \,  \Tc_{\ell' m_{\ell'} , \ell m_{\ell}}^{(u,u)}(\p,\k) \, , \nn
\end{align}
where $P = P'$ is the total energy-momentum of the three-body system and is conserved. We use the $(u,u)$ notation of Refs.~\cite{Hansen:2014eka,Hansen:2015zga}, which indicate that the amplitude is an asymmetric object, meaning we have singled out the momenta $\k$ and $\p$ as the only possible spectators. For our purposes, the asymmetric amplitude in the pair-spectator basis is the central object, however the fully symmetric $\3\to\3$ amplitude can be recovered by contracting the angular momentum indices over spherical harmonics of the pair decay angles, and summing over all possible spectators as described in Sec.~\ref{sec:scattering}. 

The $T$ matrix element contains a known singularity associated with the on-shell spectator particle which does not interact with the pair, forcing the separation
\begin{align}
	\label{eq:Tuu_connected}
	\Tc^{(u,u)}(\p,\k) \equiv \delta(\p,\k)\,\Mc_{2}(\sigma_{k}) + \Mc_{3}^{(u,u)}(\p, \k) \, ,
\end{align}
where $\Mc_2$ is the connected $\2\to\2$ amplitude of the pair, $\sigma_k \equiv (P-k)^2$ is the invariant mass squared of the pair, and $\Mc_3^{(u,u)}$ is the connected $\3\to\3$ scattering amplitude. We have also introduced the convenient shorthand $\delta(\p,\k) \equiv (2\pi)^3 \, 2\omega_k \, \delta^{(3)}(\p -\k)$ for the spectator with $\omega_k = \sqrt{m^2 + \k^2}$ being its on-shell energy. The $\2\to\2$ scattering amplitude $\Mc_2$ can be written in terms of the two-particle $K$ matrix $\Kc_2$ and the kinematic function $\Ic$ by the on-shell representation $\Mc_2^{-1} = \Kc_2^{-1} + \Ic$. We define $\Ic$ explicitly in Sec.~\ref{sec:scattering_equations}, but note here that it is a known function associated with the two-particle kinematic phase space describing on-shell pair-production.

 The analytic structure of the $T$ matrix in Eq.~\eqref{eq:Tuu_onshell} is governed by the kinematic function $\Gamma$, which is a known scheme-dependent kinematic function which has a fixed imaginary part in the physical region as imposed by unitarity. This function contains two contributions which describe the non-analyticities arising from on-shell intermediate states,
\begin{align}
	\label{eq:Gammafcn}
	\Gamma(\p,\k) \equiv \delta(\p,\k) \, H(\sigma_k) \,  \Ic(\sigma_k) + \Gc(\p,\k) \, ,
\end{align} 
where $H$ is a cutoff function introduced to render the integrals finite, $\Ic$ is the two-body kinematic function, and $\Gc$ is a known function detailing on-shell particle exchange processes between pairs, that is it switches which particle is the spectator in the intermediate state. We define these functions in detail in Sec.~\ref{sec:scattering_equations}, and stress here that these functions are not unique as only their imaginary part is fixed by the necessary non-analytic structure enforced by unitarity.

The three-body $K$ matrix is an unknown real function which contains dynamical information of the scattering process. Since $\Gamma$ is a scheme-dependent function, we require that the $K$ matrices are also scheme dependent in order to compensate the behavior to ensure that the physical scattering amplitude is scheme-independent. The scheme dependence originates from the fact that unitarity constraint only effects on-shell physics, and we can shift off-shell behavior from the kinematic function to the unknown $K$ matrix by redefining the object to absorb these effects. Compared to the two-particle case, additional scheme-dependencies come from the integration limits, which must in general be regulated to avoid ultra-violet (UV) divergences, however these play the same roles and we adopt the notion that any definition of the $K$ matrix is a scheme-dependent choice which does not affect the physical amplitude. Similar to the $T$ matrix element, the $\3\to\3$ $K$ matrix also has a disconnected structure 
\begin{align}
	\label{eq:Kmat_connected}
	\Kc_{3}^{(u,u)}(\p,\k) \equiv \delta(\p,\k) \, \Kc_{2}(\sigma_{k}) + \Kc_{3,\df}^{(u,u)}(\p,\k) \, ,
\end{align}
where the first term contains the usual $\2\to\2$ connected $K$ matrix, and the connected $\3\to\3$ $K$ matrix is denoted $\Kc_{3,\df}^{(u,u)}$, where the $\df$ stands for \emph{divergence free} which is a notation borrowed from the original RFT derivation in Ref.~\cite{Hansen:2014eka}. In Sec.~\ref{sec:alternative} we show that the connected three-body $K$ matrix is the same object first introduced in Ref.~\cite{Hansen:2014eka}. Using the aforementioned relations, we find a new integral equation for the connected $\3\to\3$ amplitude
\begin{align}
	\label{eq:M3_onshell}
	\Mc_{3}^{(u,u)}(\p,\k) & = \Kc_{3,\df}^{(u,u)}(\p,\k) - \Kc_{2}(\sigma_{p}) \, \Gc(\p,\k) \, \Mc_{2}(\sigma_{k}) \nn \\[5pt]
	& - \int_{k'} \Kc_{3,\df}^{(u,u)}(\p,\k') \, \Gamma(\k',\k) \, \Mc_{2}(\sigma_{k}) - \Kc_{2}(\sigma_{p}) \int_{k'} \Gamma(\p,\k') \, \Mc_{3}^{(u,u)}(\k',\k)   \nn \\[5pt]
	& - \int_{p'}\int_{k'} \Kc_{3,\df}^{(u,u)}(\p,\p') \, \Gamma(\p',\k') \, \Mc_{3}^{(u,u)}(\k',\k) \, .
\end{align}

As we show in Sec.~\ref{sec:alternative}, the new relation Eq.~\eqref{eq:M3_onshell} can be simply related to all other previously obtained forms, most directly that of Ref.~\cite{Hansen:2015zga}. This new representation allows for one to reconstruct the amplitude for given $K$ matrices with a single integral equation, instead of the set presented in Ref.~\cite{Hansen:2015zga}, which in turn allows for a more direct computational approach. In Sec.~\ref{sec:alternative} it is shown how to recover the original set of integral equations of Ref.~\cite{Hansen:2015zga}, where we follow a procedure of isolating all rescattering effects which do not include three-body interactions governed by $\Kc_{3,\df}^{(u,u)}$. Then, we use symmetrization identities to recover the symmetric form $\Kc_{3,\df}$ of Ref.~\cite{Hansen:2014eka}. This connection allows us to write a similar relation to Eq.~\eqref{eq:M3_onshell} in terms of the symmetric $K$ matrix, shown in Sec.~\ref{sec:symmetric_K}.

Once on-shell representations such as Eq.~\eqref{eq:M3_onshell} are constructed, we can then embed the system in a finite cubic volume of size $L$ subject to periodic boundary conditions. All momenta are quantized accordingly as 
\begin{align}
	\k = \frac{2\pi}{L}\n \, , \nn 
\end{align}
where $\n \in \Zbb^3$. Therefore, all integrals in the on-shell relations are replaced by discrete sums over quantized momenta. We find in Sec.~\ref{sec:QC} that using the FVU approach with $T$ matrix elements yields the TOPT version of the RFT quantization condition, which is given by
\begin{align}
	\label{eq:QC3_v1}
	\det \Big[ \, 1 + \left( \wt{\Kc}_{2,L} + \Kc_{3,\df}^{(u,u)}(P)  \right) \cdot \left( \wt{F}_{2,L} + \wt{G}_L(P)\right ) \,\Big]_{E = E_{\mathfrak{n}} }= 0 \, . 
\end{align}
which is a determinant in $k\ell m_{\ell}$ space where $\k$ is fixed by the momentum quantization $\k = 2\pi \n / L$. Here $\wt{F}_{2,L}$ and $\wt{G}_L$ are known functions which are finite-volume analogues of the kinematic functions $\Ic$ and $\Gc$, respectively, while $\wt{\Kc}_{2,L} = 2\omega_k L^3 \, \Kc_2(\sigma_k) \rvert_{\k = 2\pi \n /L}$. For a given total momentum $\P$, the Eq.~\eqref{eq:QC3_new} holds at the finite-volume spectrum $E = E_{\mathfrak{n}}$. Alternatively, we can look at the spectrum of the finite-volume analogue of $\Mc_3^{(u,u)}$. The reshuffling of rescattering terms cause by the disconnected two-particle subprocesses gives a different structure to the quantization condition, which we find
\begin{align}
	\label{eq:QC3_new}
	\det \Big[ \, 1 + \Kc_{3,\df}^{(u,u)}(P) \cdot F_{3,L}^{(u,u)}(P) \,\Big]_{E = E_{\mathfrak{n} } }= 0 \, ,
\end{align}
where the determinant is again over $k\ell m_{\ell}$ space and $F_{3,L}^{(u,u)}$ is a new function which depends both on the geometric nature of the finite cubic volume as well as the dynamics of the two-particle subsystem,
\begin{align}
	\label{eq:F3uu}
	F_{3,L}^{(u,u)}(P) \equiv \left[\,\wt{F}_{2,L} + \wt{G}_{L}(P)\,\right] \left( \, 1 - \frac{1}{{\wt{\Kc}_{2,L}}^{\, -1} + \wt{F}_{2,L} + \wt{G}_{L}(P)} \cdot \left[\,\wt{F}_{2,L} + \wt{G}_{L}(P)\,\right] \right) \, .
\end{align}
Equation~\eqref{eq:QC3_new} is a new form of the three-particle quantization condition in terms of the asymmetric $K$ matrix, which bears a similar structure to the one by Ref.~\cite{Hansen:2014eka} in terms of the symmetric three-body $K$ matrix. Using the same symmetrization identities on the $K$ matrices as was done for the infinite-volume scattering amplitude, we show in Sec.~\ref{sec:symmetric_K} that we can recover the original symmetric version of the three-body $K$ matrix and quantization condition.

\section{Scattering amplitudes}
\label{sec:scattering}

We consider a single scattering channel consisting of three identical spinless particles with mass $m$, \eg $3\pi^+$ elastic scattering. While the procedure works for arbitrary types of particles, \cf Refs.~\cite{Jackura:2018xnx,Mikhasenko:2019vhk}, we focus on the identical particle case in order to present various relations to all previously known results. In some general Lorentz frame~\footnote{We work in a $3+1$ Minkowski spacetime with metric signature $\mathrm{diag}(+1,-1,-1,-1)$.}, the three-particle system carries a total four-momentum $P = (E,\P)$ with energy $E$ and momentum $\P$. To suppress contributions from other multiparticle states, we assume that (\textit{i}) even-to-odd number particle transitions are forbidden by the symmetry of the underlying interaction, \eg $\2\leftrightarrow \3$ transitions are prohibited, and (\textit{ii}) the total energy is restricted to a kinematic domain such that only three-particle states can go on-shell, \ie $(3m)^2 \le s < (5m)^2$ where $s$ is the invariant-mass-squared of the three-particle system,
\begin{align}
	s \equiv P^2 = E^2 - \P^2 \, . \nn 
\end{align}
Each particle has a four-momentum $k_j = (\omega_{k_j},\k_j)$ with $j=1,2,3$ denoting the particle, and has an energy fixed by the usual relativistic dispersion relation $\omega_{k} =\sqrt{ m^2 + \k^2}$. For a fixed $P$, the momentum of the system is constrained as $P = k_1+k_2+k_3$ from momentum conservation.

We define the $T$ matrix element for $\3\to\3$ scattering, $\Tc$,~\footnote{Note that Ref.~\cite{Hansen:2015zga} and subsequent references use $\Tc$ to denote a different scattering quantity, which is discussed in Sec.~\ref{sec:alternative}, and should not be confused here with the $T$ matrix elements.} as
\begin{align}
	\label{eq:Tmatrix_element}
	\bra{\p_1,\p_2,\p_3}T\ket{\k_1,\k_2,\k_3} = (2\pi)^{4}\delta^{(4)}(P'-P)\, \Tc \, ,
\end{align}
where $\ket{\k_1,\k_2,\k_3}$ is the initial three particle state of total momentum $P$, which is defined as the symmetrized direct product over the three individual single particle states $\ket{\k_j}$~\footnote{States of three identical particles are defined by the usual tensor product
\[
  \ket{\k_1,\k_2,\k_3} \equiv \frac{1}{\sqrt{3!}}\sum_{\pi\in S_3} \ket{\k_{\pi(1)}}\otimes\ket{\k_{\pi(2)}}\otimes\ket{\k_{\pi(3)}}\, ,
\]
where $S_3$ is the permutation group of order three.}, and $\ket{\p_1,\p_2,\p_3}$ is the final state with total momentum $P'$ defined analogously. The amplitude depends on eight kinematic variables constructed from the set of initial and final state momenta $\Pc_k \equiv\{\k_1,\k_2,\k_3\}$ and $\Pc_p \equiv \{\p_1,\p_2,\p_3\}$, respectively, with total momentum conservation enforced by the Dirac delta function. Instead of specifying a set of eight variables, we show the amplitude as a function of the initial and final sets of momenta. We fix the normalization of the amplitude from the single particle states, which are chosen to be normalized in the standard relativistic way 
\begin{align}
	\braket{\p|\k} = (2\pi)^{3} \, 2\omega_{k} \delta^{(3)}(\p-\k) \equiv \delta(\p,\k) \, , \nn
\end{align}
where the latter definition will prove as a convenient shorthand. Since we consider the scattering of three identical scalars, the amplitude is a Lorentz scalar and is symmetric under interchange of the particles due to Bose statistics. Additionally, the amplitude is symmetric under time reversal symmetry since we assume we work with hadrons within QCD. Note that one can further link the scattering amplitude defined in Eq.~\eqref{eq:Tmatrix_element} to quantum correlation functions via the Lehmann-Symanzik-Zimmermann (LSZ) reduction formula. Since our goal is to work solely with on-shell amplitudes via the unitarity constraint Eq.~\eqref{eq:Tmatrix_unitarity}, it is not necessary to specify the any details of quantum fields for a generalized set of Feynman rules as is done in Refs.~\cite{Hansen:2014eka,Hansen:2015zga}.
%
\begin{figure}[t!]
    \centering
    \includegraphics[ width=0.55\textwidth]{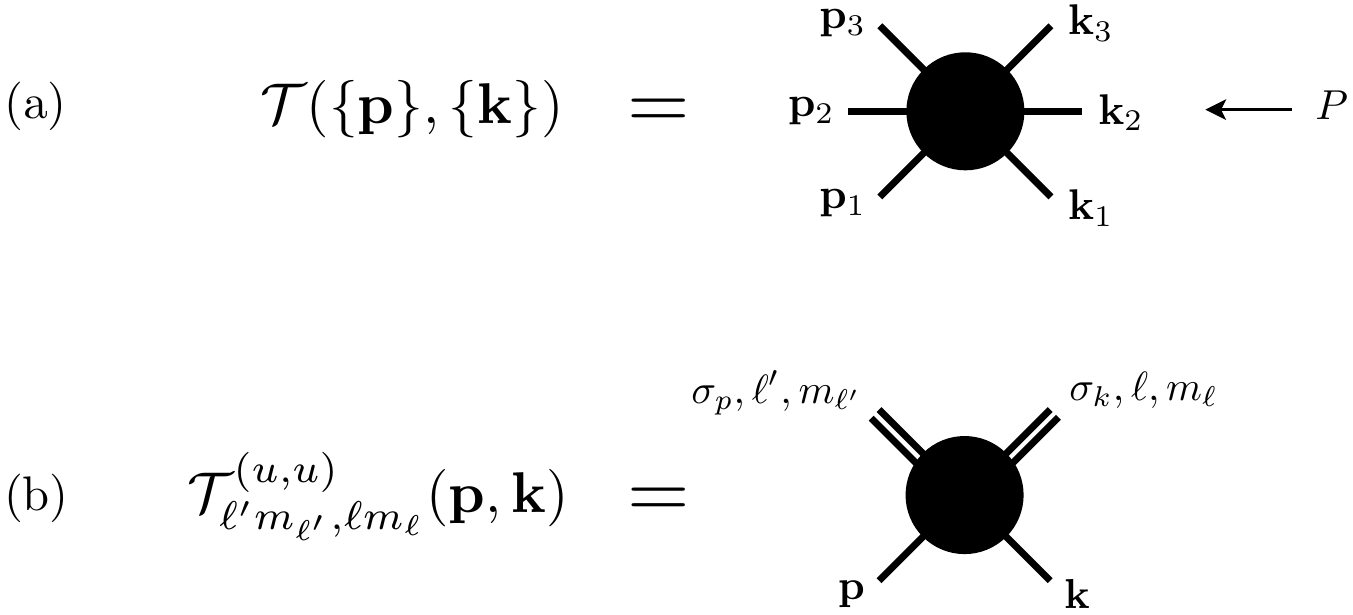}
    \caption{Scattering diagrams for (a) the $\3\to\3$ scattering amplitude defined as the $T$ matrix element in Eq.~\eqref{eq:Tmatrix_element} and (b) the pair-spectator amplitude $\Tc^{(u,u)}$ defined in \eqref{eq:Tmatrix_symm}. In these and all subsequent diagrams, momentum flow is from right to left. The single line on the amplitude in (b) indicates the spectator whereas the double line represents that pair coupled to a definite angular momentum state.}
    \label{fig:3to3_amplitudes}
\end{figure}

\begin{figure}[t!]
    \centering
    \includegraphics[ width=0.55\textwidth]{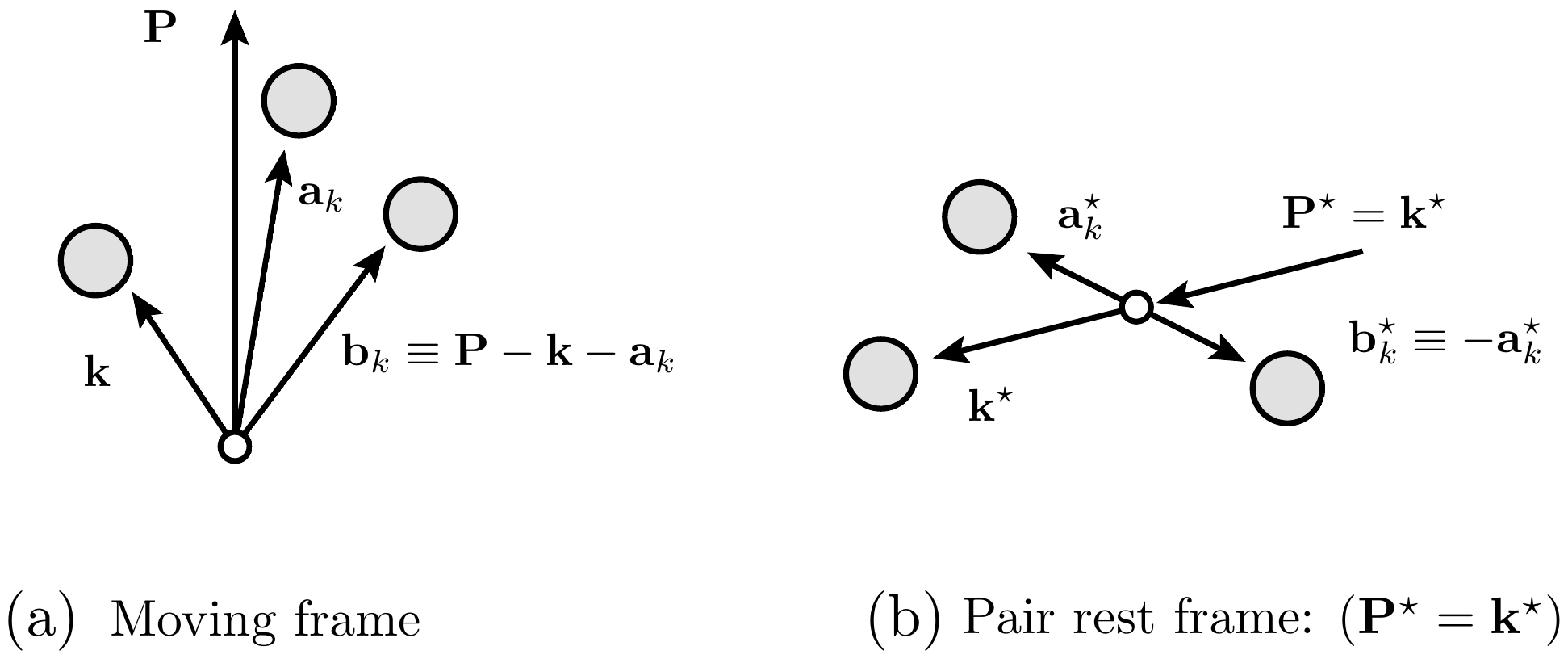}
    \caption{Kinematic assignment of a three particle state in the pair-spectator basis in (a) a general moving frame with total momentum $\P$ and (b) the pair CM frame $\P^{\star}-\k^{\star} = \0$. The angular momentum of the pair ($\ell,m_{\ell}$) is defined in its CM frame with respect to the angles of $\a_k^{\star}$, which has a magnitude fixed by the pair invariant mass $\lvert\a_k^{\star}\rvert = q_{k}^{\star}$ as described in the text.}
    \label{fig:3B_kinematics}
\end{figure}

To simplify the degrees of freedom, it is convenient to work in a pair-spectator basis, defined such that two of the particles couple in a \textit{pair}~\footnote{The pair associated with the spectator is also called an isobar or dimer in the literature.} with definite angular momentum $\ell$ and projection $m_{\ell}$ which scatters against the third particle which is called the \textit{spectator}. We define such a state by singling out one particle, \eg particle $1$, to be the spectator and carry a momentum $\k$. The other two particle constitute the pair, \ie particles $2$ and $3$, in which we choose a cyclic convention to fix the first particle of the pair, \ie $2$ is the first particle in the pair. We label the momentum of the first particle in the pair as $\a_k$, and the second particle has a momentum $\b_k \equiv \P - \k - \a_k$ by momentum conservation. The subscripts $k$ serve to inform that this pair belongs to the spectator with momentum $\k$. We now Lorentz boost 
to the pair center-of-momentum (CM) frame, defined when $\P^{\star}-\k^{\star} = \0$, where the $\star$ indicates that the kinematic variables are taken in this frame. Figure~\ref{fig:3B_kinematics} shows a cartoon of this reference frame. Then, $\a_k^{\star} = -\b_k^{\star} \equiv q_k^{\star}\,\bh{\a}_k^{\star}$, where $q_k^{\star} \equiv \lvert \a_k^{\star} \rvert = \sqrt{\sigma_k/4-m^2}$ is the relative momentum of the pair which is fixed by its invariant mass and $\bh{\a}_k^{\star}$ is the orientation of the first particle, which is not fixed. We can then introduce a partial wave expansion on this state, and use the angle $\bh{\a}_k^{\star}$ to define the angular momentum of the pair. A three particle state $\ket{\k_1,\k_2,\k_3}$ has three spectators, which we enforce by summing over all permutations of the previous steps, leading the to relation
\begin{align}
	\label{eq:state}
	\ket{\k_1 , \k_2, \k_3} = \sqrt{4\pi} \, \sum_{\k \in \Pc_k} \sum_{\ell,m_{\ell}} \ket{P, \k ,\ell m_{\ell}} Y_{\ell m_{\ell}}^{*}(\bh{\a}_{k}^{\star} ) \, ,
\end{align}
where the sum is over the set of momenta $\Pc_k = \{\k_1,\k_2,\k_3\}$ used to recover all spectator contributions. The state $\ket{P,\k,\ell m_{\ell}}$ describes a system of spectator with momentum $k=(\omega_k,\k)$ recoiling against a pair with momentum 
\begin{align}
	P_k \equiv P-k = (E-\omega_k,\P-\k) \, , \nn 
\end{align}
which has an invariant-mass-squared
\begin{align}
	\sigma_k \equiv P_k^2 = (P-k)^2 = (E-\omega_k)^2 - (\P-\k)^2 \, . \nn
\end{align}
In a physical scattering process, the pair invariant mass can take values $(2m)^2 \le \sigma_{k} \le (\sqrt{s} - m)^2$. Bose symmetry imposes that states with only even $\ell$ are allowed, hence the two particle state in the pair CM frame has positive parity.
%

\begin{figure}[t!]
    \centering
    \includegraphics[ width=0.70\textwidth]{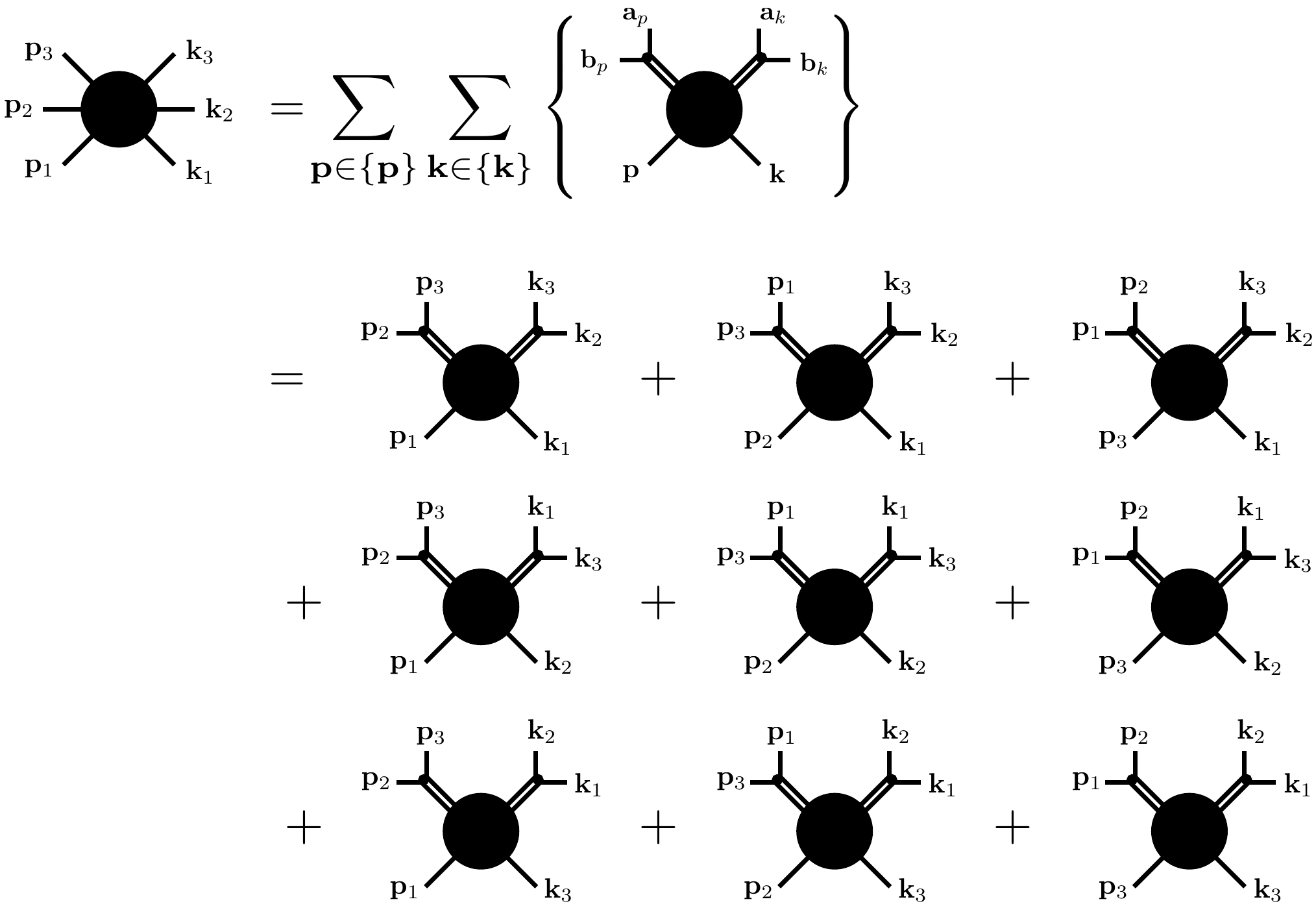}
    \caption{The $\3\to\3$ amplitude expressed as a sum over pair-spectator amplitudes as defined in Eq.~\eqref{eq:Tmatrix_symm}. The vertices connecting the two external lines to a double line represent the angular couplings to angular momenta via spherical harmonics.}
    \label{fig:3B_sum_over_spectators}
\end{figure}

We use Eq.~\eqref{eq:state} for both the initial and final states to rewrite the symmetric $\3\to\3$ amplitude in the pair-spectator basis,
\begin{align}
	\label{eq:Tmatrix_symm}
	\Tc = \sum_{\p\in \Pc_p} \sum_{\k \in \Pc_k} \bigg\{ \, 4\pi \sum_{ \ell',m_{\ell'} } \sum_{ \ell, m_{\ell} } Y_{ \ell' m_{\ell'} }(\bh{\a}_{p}^{\star}) \,  \Tc_{\ell' m_{\ell'} , \ell m_{\ell} }^{(u,u)} (\p,\k)  \, Y_{ \ell m_{\ell} }^{*}(\bh{\a}_{k}^{\star}) \, \bigg\} \, ,
\end{align}
which separates the full amplitude into sum over nine pair-spectator amplitudes $\Tc^{(u,u)}$, as shown in Fig.~\ref{fig:3B_sum_over_spectators}, which are defined as the $T$ matrix elements
\begin{align}
	\bra{P',\p,\ell' m_{\ell'}} T \ket{P,\k,\ell m_{\ell}} = (2\pi)^{4}\delta^{(4)}(P' -P) \,  \Tc_{\ell' m_{\ell'} , \ell m_{\ell}}^{(u,u)}(\p,\k) \, . \nn
\end{align}
Here we adopted the notation introduced in Ref.~\cite{Hansen:2014eka} that the superscripts $(u,u)$ indicate that this is an asymmetric amplitude specified by spectators with momentum $\k$ and $\p$ for the initial and final states, respectively.~\footnote{This has been called an isobar expansion, for instance in Ref.~\cite{Mai:2017vot,Jackura:2018xnx,Dawid:2020uhn}.} Four of the eight kinematic variables are given by $\bh{\a}_k^{\star}$ and $\bh{\a}_p^{\star}$, leaving four remaining degrees of freedom to describe $\Tc^{(u,u)}$. One of those variables is the total invariant mass $s$, or equivalently the total momentum $P$, which we suppress while the other three can be formed from the two spectator momenta in an appropriate frame. The amplitude $\Tc^{(u,u)}$ is similar to an effective coupled $\2\to\2$ process of a spinless particle scattering against a ``particle'' with spin $\ell$ and mass $\sqrt{\sigma_k}$, which transitions to another ``particle'' with spin $\ell'$ and mass $\sqrt{\sigma_p}$, the primary exception being that the masses are variable.
%

\begin{figure}[t!]
    \centering
    \includegraphics[ width=0.60\textwidth]{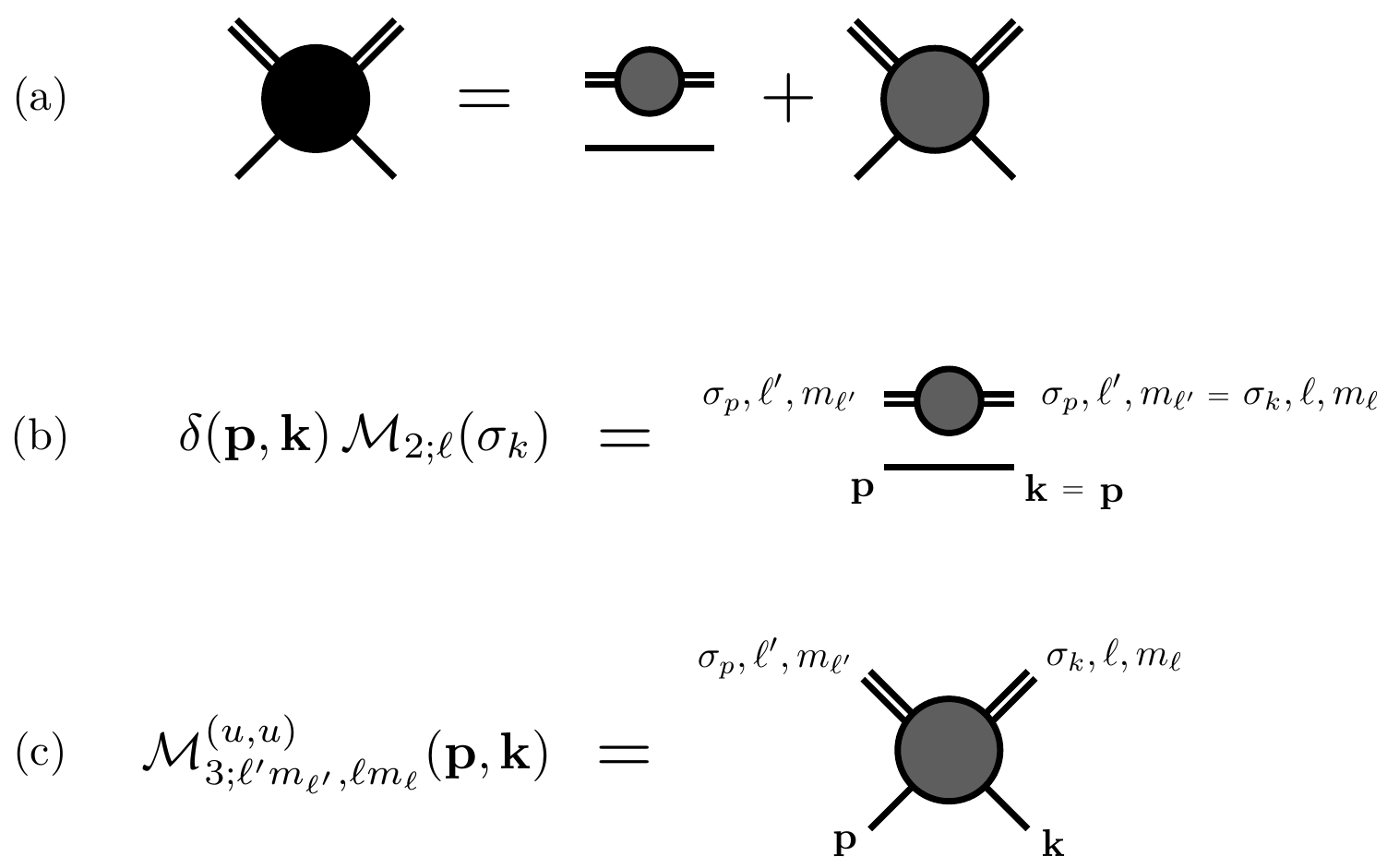}
    \caption{(a) The cluster decomposition of the $T$ matrix element Eq.~\eqref{eq:Tuu_connected} into (b) the disconnected contribution involving the connected $\2\to\2$ scattering amplitude $\Mc_2$, and (c) the connected $\3\to\3$ amplitude $\Mc_3^{(u,u)}$.}
    \label{fig:amplitudes_conn}
\end{figure}

The $T$ matrix elements for $\3\to\3$ processes contain two causally disconnected structures according to the cluster decomposition principle.~\footnote{The $\3\to\3$ $S$ matrix contains $3!=6$ fully disconnected contributions and $3\times 3 = 9$ disconnected contributions when one of the particles remains causally separated~\cite{Weinberg:1995mt}. Since we choose to work with pair-spectator amplitudes, Eq.~\eqref{eq:Tmatrix_symm}, the nine combinations of spectator momenta agrees with that of the partially disconnected terms, giving us one term containing $\2\to\2$ amplitude in Eq.~\eqref{eq:Tuu_connected}.} The first is where two of the particles undergo interactions while the third remains a non-interacting spectator. The second term is the fully connected $\3\to\3$ amplitude where there are interactions amongst all three particles. For the pair-spectator amplitude $\Tc^{(u,u)}$, the cluster decomposition principle imposes the structure Eq.~\eqref{eq:Tuu_connected}, which we repeat here
\begin{align}
	\Tc^{(u,u)}(\p,\k) = \delta(\p,\k)\,\Mc_{2}(\sigma_{k}) + \Mc_{3}^{(u,u)}(\p, \k) \, , \nn 
\end{align}
where $\Mc_2$ is the $\2\to\2$ partial wave amplitude, $\delta(\p,\k)$ conserves the spectator momentum, and $\Mc_{3}^{(u,u)}$ is the connected $\3\to\3$ pair-spectator amplitude, \cf Fig.~\ref{fig:amplitudes_conn} for a diagram of this decomposition. If one desires, a complete $\3\to\3$ connected scattering amplitude $\Mc_3$ can be constructed by summing over all pair-spectator combinations, yielding a similar relation as Eq.~\eqref{eq:Tmatrix_symm}. The $\2\to\2$ partial wave amplitude is the same as the usually defined $T$ matrix element of the two-hadron angular momentum state with a total momentum $P_k$,
\begin{align}
	\bra{P_p,\ell' m_{\ell'}}T\ket{P_k,\ell m_{\ell}} = (2\pi)^4\delta^{(4)}(P_p - P_k) \, \Mc_{2;\ell' m_{\ell'} , \ell m_{\ell}}(\sigma_k) \, . \nn 
\end{align}
The $\2\to\2$ amplitude is a diagonal matrix in angular momentum space and independent of the projection $m_{\ell}$ due to rotational invariance, $\Mc_{2;\ell' m_{\ell'} , \ell m_{\ell} } = \delta_{\ell' \ell} \delta_{m_{\ell'} m_{\ell}} \, \Mc_{2;\ell}$. Furthermore, it depends only on a single degree of freedom, which we choose as the invariant $\sigma_k$. For a fixed physical $(3m)^2 \le s < (5m)^2$, the $\2\to\2$ amplitude lies in the kinematic region $(2m)^2 \le \sigma_k < (4m)^2$, which is precisely the physical domain which closes off higher multi-particle states.
%
%

\subsection{Unitarity relations}
\label{sec:unitarity}

\begin{figure}[t!]
    \centering
    \includegraphics[ width=0.60\textwidth]{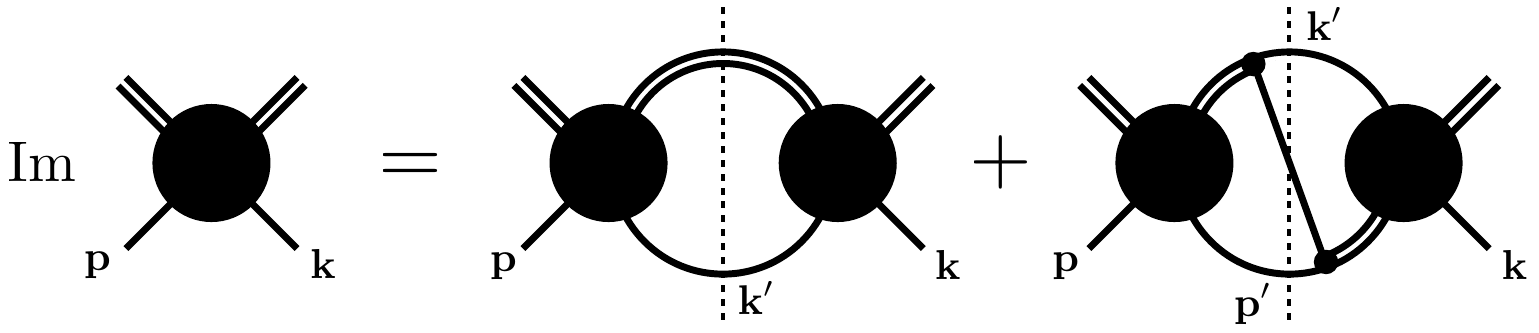}
    \caption{Unitarity relation given in Eq.~\eqref{eq:Tuu_unitarity}. The vertical dashed lines indicate the direct (first term) and exchange (second term) cuts corresponding to the functions $\rho(\sigma_{k'})$ and $\Delta(\p',\k')$, respectively. Each closed loop represents a three-dimensional integral over the on-shell spectator momentum, and we use the convention that the amplitude on the left of the dashed line is conjugated.}
    \label{fig:3to3.unitarity.A3_unitarity}
\end{figure}

Following the procedure discussed in Refs.~\cite{Mai:2017vot,Jackura:2018xnx}, we form matrix elements of Eq.~\eqref{eq:Tmatrix_unitarity} using the basis defined by Eq.~\eqref{eq:state} to obtain the unitarity relation for the pair-spectator amplitude. Since energies are such that only three particle states are on-shell, we need only consider the unitarity condition for elastic $\3\to\3$ scattering. Therefore, the unitarity relation for the amplitude $\Tc^{(u,u)}$ is given by
\begin{align}
	\label{eq:Tuu_unitarity}
	\im \Tc^{(u,u)} (\p,\k) & = \int_{k'} \, \Tc^{(u,u)\,\dag }(\p,\k')  \, \rho(\sigma_{k'}) \,  \, \Tc^{(u,u)}(\k',\k)  \nn \\[5pt]
	& +  \int_{p'} \int_{k'} \, \Tc^{(u,u)\,\dag }(\p,\p')   \, \Delta(\p',\k') \,   \Tc^{(u,u)}(\k',\k) \, ,
\end{align}
which is an infinite dimensional matrix equation in angular momentum space where the $\ell$ and $m_{\ell}$ indices are suppressed. For convenience, we present the derivation of Eq.~\eqref{eq:Tuu_unitarity} is given in Appendix~\ref{sec:app_unitarity}. Two kinematic functions are introduced in Eq.~\eqref{eq:Tuu_unitarity}, the first being the usual two-body phase space $\rho$, given by
\begin{align}
	\label{eq:direct_cut}
	\rho_{\ell' m_{\ell'} , \ell m_{\ell}} (\sigma_{k}) & = \delta_{\ell' \ell} \delta_{m_{\ell'} m_{\ell}} \,  \frac{\xi q_{k}^{\star}}{8\pi \sqrt{\sigma_{k}} } \, \Theta( q_{k}^{\star}) \, , 
\end{align}
where $\xi = 1/2!$ is the symmetry factor for two identical particles. The Heaviside function $\Theta(q_k^{\star})$ enforces that the intermediate state exists when the pair CM energy allows for on-shell pair production, which induces a square root branch cut in the $\sigma_k$-plane when $\sqrt{\sigma_k} \ge 2m$. We call the first term of Eq.~\eqref{eq:Tuu_unitarity} the \emph{direct} term since the spectator remains unchanged in the intermediate state. The second kinematic function is the single particle exchange pole, $\Delta$, defined as
\begin{align}
	\label{eq:exchange_cut}
	\Delta_{\ell' m_{\ell'} ,  \ell m_{\ell}}(\p,\k)   = \pi \, & \delta\left( (P - k - p)^2 - m^2 \right)  \, \nn \\[5pt]
	& \times  4\pi Y_{\ell' m_{\ell'}}^{*}(\bh{\k}_{p}^{\star}) Y_{\ell m_{\ell}}(\bh{\p}_{k}^{\star}) \Big\rvert_{(P - k - p)^2 = m^2 } \,  \Theta\left( q_p^{\star} \right) \Theta\left( q_k^{\star} \right) \, ,
\end{align}
which characterizes the singularity associated with exchanging an on-shell particle between pairs.~\footnote{The phases of the spherical harmonics cancel each other in the unitarity condition as it sums over all possible intermediate states. One can avoid keeping tack of all phase cancellations in intermediate steps by considering real spherical harmonics, which respect the same completeness and orthogonality relations as the standard one.} The second term of Eq.~\eqref{eq:Tuu_unitarity} will be referred to as the \emph{exchange} or \emph{recoupling} term due to the switching of spectators in the intermediate state. The angles associated with a spectators in each spherical harmonic are defined with respect to the pair associated with the opposite spectators CM frame. For example, $\bh{\k}^{\star}_p$ is the orientation of the initial state pair in the CM frame of the final state pair associated with the spectator $p$. The two Heaviside functions enforce that the pair is on-shell, while the Dirac delta function fixes the exchange particle to its mass-shell. Note that the Heaviside functions limit the kinematics in the argument of the delta function to forward propagation only, however we keep the covariant representation for convenience. Figure~\ref{fig:3to3.unitarity.A3_unitarity} shows a diagrammatic representation of Eq.~\eqref{eq:Tuu_unitarity}, illustrating the the direct and exchange processes which cause the non-analyticities in Eqs.~\eqref{eq:direct_cut} and \eqref{eq:exchange_cut}, respectively. Equation~\eqref{eq:Tuu_unitarity} is a non-perturbative statement for the on-shell amplitude for any interaction satisfying the  assumptions on particle content and allowed kinematic domain discussed at the beginning of Sec.~\ref{sec:scattering}.
%

\begin{figure}[t!]
    \centering
    \includegraphics[ width=0.95\textwidth]{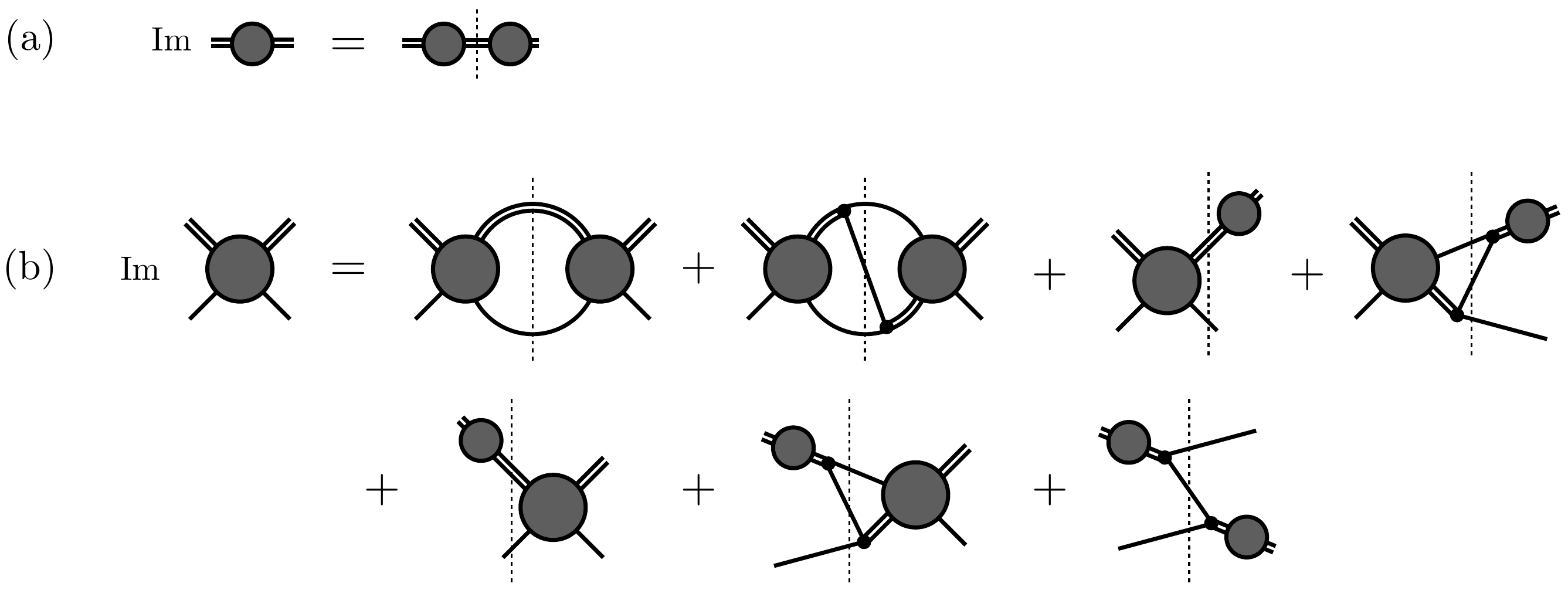}
    \caption{Diagrammatic representation of the connected unitarity conditions for the (a) $\2\to\2$ amplitude Eq.~\eqref{eq:M2_unitarity}, and (b) the $\3\to\3$ amplitude Eq.~\eqref{eq:M3_unitarity}.}
    \label{fig:unitarity_conn}
\end{figure}

In addition to unitarity for $T$ matrix elements, we can introduce separate unitarity conditions for the connected amplitudes. Inserting Eq.~\eqref{eq:Tuu_connected} into \eqref{eq:Tuu_unitarity} results in the unitarity relation separating into two constraints, 
\begin{align}
	\label{eq:M2_unitarity}
	\im \Mc_2(\sigma_{k}) = \Mc_2^{\dag}(\sigma_{k}) \, \rho(\sigma_{k}) \, \Mc_2(\sigma_{k}) \, ,
\end{align}
which is the usual condition for the $\2\to\2$ partial wave amplitude, and 
\begin{align}
	\label{eq:M3_unitarity}
	\im \Mc_{3}^{(u,u)} (\p,\k) & =  \int_{p'} \int_{k'} \, \Mc_{3}^{(u,u)\,\dag }(\p,\p')   \, \Phi(\p',\k') \,   \Mc_{3}^{(u,u)}(\k',\k) \nn \\[5pt]
	& + \int_{k'}  \, \Mc_{2}^{\dag}(\sigma_{p}) \,  \Phi(\p,\k')  \, \Mc_{3}^{(u,u)}(\k',\k) \nn \\[5pt]
	& +  \int_{p'} \, \Mc_{3}^{(u,u)\,\dag }(\p,\p')  \, \Phi(\p',\k)  \, \Mc_{2}(\sigma_{k}) \nn \\[5pt]
	& + \Mc_{2}^{\dag} (\sigma_{p}) \,  \Delta(\p,\k) \, \Mc_{2}(\sigma_{k}) \, ,
\end{align}
which is the condition for the connected $\3\to\3$ amplitude. Here we have introduced a convenient shorthand, the matrix $\Phi$, which contains both the direct and exchange kinematic functions
\begin{align}
	\label{eq:Phi_cut}
	\Phi(\p,\k) \equiv \delta(\p,\k) \, \rho(\sigma_{k}) + \Delta(\p,\k) \, .
\end{align}
Figure~\ref{fig:unitarity_conn} presents a diagrammatic picture of the connected unitarity relations. The first term of Eq.~\eqref{eq:M3_unitarity} is analogous to the $\2\to\2$ case with the addition of intermediate states particle exchanges, while the other three terms are associated with rescattering effects of the two particle sub-channels. The last term of Eq.~\eqref{eq:M3_unitarity} represents the pole singularity associated with on-shell particle exchange between two $\2\to\2$ sub-processes. Ultimately we aim to describe the connected $\Mc_2$ and $\Mc_3$ amplitudes by constructing suitable on-shell representations satisfying Eqs.~\eqref{eq:M2_unitarity} and \eqref{eq:M3_unitarity}, respectively, as in \eg Refs.~\cite{Mai:2017vot,Jackura:2018xnx,Mikhasenko:2019vhk}. Instead, in the following section we work directly with Eq.~\eqref{eq:Tuu_unitarity} to find an on-shell representation, and then use Eq.~\eqref{eq:Tuu_connected} to find an expression for the $\Mc_3$ amplitude. This direction proves to be a simpler, more direct route toward obtaining on-shell representations for $\3\to\3$ amplitude, and builds on the intuition learned from $\2\to\2$ scattering. In Section~\ref{sec:alternative}, we present alternative approaches which work with the connected unitarity condition, and show the relations between the results found in Section~\ref{sec:unitarity}.
%
%

\subsection{On-shell scattering equations}
\label{sec:scattering_equations}

Nonlinear constraints such as Eq.~\eqref{eq:Tuu_unitarity} are solved by linear equations that are driven by a kernel, known as a $K$ matrix, which is a real function within the restricted kinematic region which we consider and represents the short-distance dynamics of the scattering process which are not fixed by the unitarity condition. As an example, we first review the well-known on-shell representation for the $\2\to\2$ amplitude that satisfies Eq.~\eqref{eq:M2_unitarity}. We make an \ansatz for the on-shell amplitude in the linear form
\begin{align}
	\label{eq:M2_onshell}
	\Mc_{2}(\sigma_{k}) = \Kc_{2}(\sigma_{k}) - \Kc_{2}(\sigma_{k}) \, \Ic(\sigma_{k}) \, \Mc_{2}(\sigma_{k}) \, ,
\end{align}
where $\Kc_2$ is the $\2\to\2$ partial wave $K$ matrix, which is an unknown real function in the elastic kinematic domain, and $\Ic$ is a kinematic function which must fulfill the constraint 
\begin{align}
	\im \Ic(\sigma_k) = -\rho(\sigma_k) \nn
\end{align}
in order to satisfy Eq.~\eqref{eq:M2_unitarity}. It is straightforward to show that Eq.~\eqref{eq:M2_onshell} satisfies unitarity~\eqref{eq:M2_unitarity}. The $K$ matrix inherits the angular momentum structure of $\Mc_2$, reducing Eq.~\eqref{eq:M2_onshell} to a single algebraic relation. Subject to these restrictions, one can easily verify that Eq.~\eqref{eq:M2_onshell} satisfies Eq.~\eqref{eq:M2_unitarity}, and therefore a solution for a given particular $\Kc_2$ constitutes an on-shell representation for the $\2\to\2$ scattering amplitude valid for the kinematic domain $(2m)^2 \le \sigma_k \le (\sqrt{s} - m)^2$. 

On-shell representations such as Eq.~\eqref{eq:M2_onshell} have an intrinsic \emph{scheme dependence} since the choice of $\Ic$, and hence the $K$ matrix, is not unique. For any $\Ic$ satisfying the requisite $\im \Ic = -\rho$, the function $\Kc_2$ must compensate the difference in order to ensure $\Mc_{2}$ is the physical scattering amplitude. Choosing an alternative $\Ic^{\,\prime}$ results in an associated $\Kc_{2}'$, which can be related to $\Kc$ through the invariance of $\Mc_2$ as 
\begin{align}
	{\Kc_2^{\,\prime}}^{-1} = {\Kc_2}^{-1} + \Ic - \Ic^{\,\prime} \, . \nn
\end{align}
Note that the relation between $K$ matrices preserves that condition that $\Kc_2$ is real in the restricted kinematic since we require $\im\Ic = \im\Ic'$. We will use two different choices of $\Ic$ in this work. The first one is trivially related to the simple phase space factor Eq.~\eqref{eq:direct_cut}
\begin{align}
	\label{eq:Ifcn}
	\Ic_{\ell' m_{\ell' } , \ell m_{\ell}} (\sigma_{k})  = -i  \delta_{\ell' \ell} \delta_{m_{\ell'} m_{\ell}} \, \frac{\xi \, q_{k}^{\star} }{8\pi \sqrt{\sigma_k} } \, ,
\end{align}
With an associated $K$ matrix $\Kc_2$. Note that this term does not depend on any explicit parameter that indicates a scheme dependence, however it is still a choice that we can make which fixes the on-shell form for $\Kc_2$. The second choice is given by an integral form
\begin{align}
	\label{eq:Iprimefcn}
	\Ic_{\ell' m_{\ell' } , \ell m_{\ell}}^{\,\prime}(\sigma_{k}) & = \int^{\Lambda} \! \frac{\diff \sigma'}{\pi} \, \frac{\rho_{\ell' m_{\ell' } , \ell m_{\ell}}(\sigma')}{\sigma_{k} -\sigma' + i\epsilon} \left(\,  \sqrt{ \frac{\sigma'-4m^2}{\sigma_{k}-4m^2} } \,\right)^{\ell'+\ell} \, ,
\end{align}
with the associated $K$ matrix $\Kc_2'$. This is the usual Chew-Mandelstam dispersive form~\cite{Chew:1960iv} including barrier factors to ensure the kinematic singularity induced by the partial wave projection is removed. It is understood that $\epsilon \to 0^+$, and the dispersive integral runs from $4m^2 \le \sigma' < \infty$, with $\Lambda$ indicating the need for regularization of the divergent behavior.~\footnote{In phenomenological analyses, it is usual to perform subtractions to tame the UV divergence.} We do not make any specific choice for UV regulation here as any choice which respects the kinematic requirements is valid for our arguments. This shows an explicit scheme dependence in $\Kc_2^{\,\prime}$, as it must depend on the regularization $\Lambda$ in order to ensure that $\Mc_2$ is regularization independent. By inspection, both Eqs.~\eqref{eq:Ifcn} and \eqref{eq:Iprimefcn} have $\im \Ic = \im \Ic' = -\rho$ above the two-body threshold, and the relation between the two $K$ matrices is 
\begin{align}
	{\Kc_2^{\,\prime}}^{-1} = {\Kc_2}^{-1} - \re \Ic^{\,\prime} \, . \nn 
\end{align}
The advantage of Eq.~\eqref{eq:Iprimefcn} is it ensures that non-physical singularities of Eq.~\eqref{eq:Ifcn}, at $\sigma_k = 0$, are removed. While the dispersive form has much better analytic properties, it is common to use both forms of $\Ic$ in phenomenological and lattice QCD studies~\cite{Briceno:2017max}. In the next section, we will use the dispersive form to form the finite-volume analogue.

We follow a similar path as the $\2\to\2$ case for constructing an on-shell representation for $\Tc$. Given the unitarity relation Eq.~\eqref{eq:Tuu_unitarity}, we make an \ansatz for the $\3\to\3$ $T$ matrix element in the form of a linear equation such that~\eqref{eq:Tuu_unitarity} is satisfied. Since Eq.~\eqref{eq:Tuu_unitarity} is a matrix in angular momentum space and contains integrals over the spectator momenta, the linear equation will take the form of a matrix-integral equation shown in Eq.~\eqref{eq:Tuu_onshell}, which we repeat here
\begin{align}
	\Tc^{(u,u)}(\p,\k) & = \Kc_{3}^{(u,u)}(\p,\k) - \int_{p'} \int_{k'} \Kc_{3}^{(u,u)}(\p,\p') \,  \Gamma(\p',\k')  \, \Tc^{(u,u)}(\k',\k) \, . \nn
\end{align}
Here, $\Kc_{3}$ is the $\3\to\3$ $K$ matrix, which represents the unknown short-distance scattering physics, and $\Gamma$ is a kinematic function containing the non-analytic structures that satisfy $\im \Gamma(\p,\k) = -\Phi(\p,\k)$ in order to respect Eq.~\eqref{eq:Tuu_unitarity}. We choose to define $\Gamma$ as in Eq.~\eqref{eq:Gammafcn}, which we repeat here for convenience
\begin{align}
	\Gamma(\p,\k) = \delta(\p,\k) \, H(\sigma_k) \, \Ic(\sigma_{k}) + \Gc(\p,\k) \, , \nn
\end{align}
where $\Ic$ is as in Eq.~\eqref{eq:Ifcn} which describes the direct production of three on-shell intermediate states, $H$ is a cutoff function to be described shortly, and $\Gc$ is the exchange propagator
\begin{align}
	\label{eq:Gfcn}
	\Gc_{\ell' m_{\ell'} , \ell m_{\ell}}(\p,\k)   =    \frac{H(\sigma_p)H(\sigma_k)}{(P - k - p)^2 - m^2 + i\epsilon}  \,  \left( \frac{k_p^{\star}}{q_p^{\star}} \right)^{\ell'} 4\pi \, Y_{\ell' m_{\ell'}}^{*}(\bh{\k}_{p}^{\star}) Y_{\ell m_{\ell}}(\bh{\p}_{k}^{\star}) \left( \frac{p_k^{\star}}{q_{k}^{\star}} \right)^{\ell} \, ,
\end{align}
which satisfies the relation $\im \Gc(\p,\k) = -\Delta(\p,\k)$ to capture the exchange singularity. The exchange function contains barrier suppression factors, which are included to ensure regularity near zero spectator momenta, and evaluate to unity at the on-shell point $(P - k - p)^2 = m^2$. Like $\Ic$, $\Gc$ is not unique as any real function in the kinematic domain can be added so long as $\im \Gc = -\Delta$ in the physical region, \eg a non-covariant form as is used in Ref.~\cite{Hansen:2014eka}. Proving that $T$ matrix unitarity~\eqref{eq:Tuu_unitarity} holds for \eqref{eq:Tuu_onshell} has been more challenging than its $\2\to\2$ counterpart, \cf Refs.~\cite{Jackura:2018xnx,Briceno:2019muc}, however we provide a simple proof in Appendix~\ref{sec:app_proof}.

In Eqs.~\eqref{eq:Gammafcn} and \eqref{eq:Gfcn}, we have introduced a general cutoff function $H$ to regulate potential UV divergences in the spectator momentum integrals. Generally, $H$ must satisfy the condition that $H = 1$ for $q_k^{\star} \ge 0$, and we choose it to zero at some kinematic point below the two-particle threshold. The minimal choice one can make is $H(\sigma_k) = \Theta(q_k^{\star})$, as was used in Ref.~\cite{Jackura:2018xnx}. Although this choice is valid with respect to the unitarity condition, Ref.~\cite{Hansen:2014eka} advocates for a smoother cutoff to avoid spurious finite volume effects that may arise, \eg Eq. (29) of Ref.~\cite{Hansen:2014eka}. While this is not easily seen from the FVU viewpoint, one can use a smooth cutoff function so long as it is unity above the two-particle threshold and the integration does not overlap with any below threshold cuts. Any different choice of the cutoff functions amounts to choosing a different definition of the $K$ matrices, effectively being part of the chosen scheme dependence. In this work, we do not choose any specific cutoff, and leave any further discussion on the validity between smooth and hard cutoffs to future work.

The $K$ matrix inherits the same connectedness structure as $\Tc^{(u,u)}$, \cf Eq.~\eqref{eq:Tuu_connected}. Splitting this function according to cluster decomposition, we define $\Kc_{3,\df}^{(u,u)}$~\footnote{In the TOPT derivations of Ref.~\cite{Blanton:2020gha}, they find an expression involving this same $K$ matrix. However, the authors take two different approaches to constructing the $K$ matrix form, and stress that some properties of the resulting $K$ matrices are different, even though the on-shell form are the same. From the point of view of $S$ matrix unitarity, if the on-shell forms are identical, that is some scheme is taken to establish the kinematic function for intermediate states, then the $K$ matrices which encapsulate \emph{all} remaining short-distance physics must be the same.} as the fully connected three particle $K$ matrix which does not contain the disconnected contribution specified by $\Kc_2$, \ie
\begin{align}
	\Kc_{3}^{(u,u)}(\p,\k) \equiv \delta(\p,\k) \, \Kc_{2}(\sigma_{k}) + \Kc_{3,\df}^{(u,u)}(\p,\k) \, , \nn 
\end{align}
which was shown in Eq.~\eqref{eq:Kmat_connected}. Both $K$ matrices are real symmetric matrices in the kinematic domain of interest. We use the notation as in Refs.~\cite{Hansen:2014eka,Hansen:2015zga,Blanton:2020gha} where the connected three-body $K$ matrix is indicated by $\df$, meaning divergent free, which indicates it is free of any kinematic singularities. Since we build in cutoff functions in the $\3\to\3$ $T$ matrix element, the two-body $K$ matrix implicitly has a dependence given by $\Mc_{2}^{-1}(\sigma_k) = \Kc_{2}^{-1}(\sigma_k) + H(\sigma_k)\,\Ic(\sigma_k)$. Although this slightly modifies the original ``unprimed'' form in Eqs.~\eqref{eq:M2_onshell} and \eqref{eq:Ifcn} by 
\begin{align}
	\Kc_{2}^{-1}(\sigma_k) \to \Kc_2^{-1}(\sigma_k) + [1 -  H(\sigma_k) ] \, \Ic(\sigma_k) \, , \nn
\end{align}
we use the same notation $\Kc_2$ to avoid introducing yet another definition.

We use Eqs.~\eqref{eq:Tuu_connected} and~\eqref{eq:Kmat_connected} to separate Eq.~\eqref{eq:Tuu_onshell} into disconnected and connected contributions, giving two relations: one equation for $\Mc_2$ given by Eq.~\eqref{eq:M2_onshell}, and an another for $\Mc_3^{(u,u)}$ given by Eq.~\eqref{eq:M3_onshell}, repeated here for convenience
\begin{align}
	\Mc_{3}^{(u,u)}(\p,\k) & = \Kc_{3,\df}^{(u,u)}(\p,\k) - \Kc_{2}(\sigma_{p}) \, \Gc(\p,\k) \, \Mc_{2}(\sigma_{k}) \nn \\[5pt]
	& - \int_{k'} \Kc_{3,\df}^{(u,u)}(\p,\k') \, \Gamma(\k',\k) \, \Mc_{2}(\sigma_{k}) - \Kc_{2}(\sigma_{p}) \int_{k'} \Gamma(\p,\k') \, \Mc_{3}^{(u,u)}(\k',\k)   \nn \\[5pt]
	& - \int_{p'}\int_{k'} \Kc_{3,\df}^{(u,u)}(\p,\p') \, \Gamma(\p',\k') \, \Mc_{3}^{(u,u)}(\k',\k) \, . \nn 
\end{align}
Given a $\Kc_{2}$ and $\Kc_{3,\df}$, Eq.~\eqref{eq:M3_onshell} can be solved for the $\3\to\3$ scattering amplitude. In the next section we construct the quantization conditions which link the $K$ matrices to finite-volume spectra of particles in a box. Note that compared with Refs.~\cite{Hansen:2015zga,Blanton:2020gha}, we have a single integral for the connected $\3\to\3$ amplitude, instead of two. We show in Sec.~\ref{sec:alternative} an alternative derivation of the connected $\3\to\3$ amplitude which allows us to recover the original set of equations shown in Refs.~\cite{Hansen:2015zga,Blanton:2020gha} and connects to Eq.~\eqref{eq:M3_onshell}.

\section{Quantization conditions}
\label{sec:QC}

We now construct the quantization conditions associated with the on-shell scattering equations. The scattering equations provide the basis for the structure of the finite-volume counterparts to the scattering amplitudes. The system is put into a cubic volume with a side length $L$ and subject to periodic boundary conditions. Consequently, rotational invariance is broken and all momentum are quantized according to 
\begin{align}
	\k = \frac{2\pi}{L} \, \n  \, ,  \nn
\end{align}
where $\n \in \Zbb^{3}$. Therefore, all momentum integrals are replaced by sums by the rule,
\begin{align}
	\int \! \frac{\diff^3\k}{(2\pi)^{3}} \longrightarrow \frac{1}{L^{3}} \sum_{\k \in (2\pi/L) \Zbb^{3}} \, . \nn 
\end{align}
Likewise, the Dirac delta functions will be replaced by the Kronecker equivalent,
\begin{align}
	(2\pi)^3 \, \delta^{(3)}(\p - \k) \longrightarrow L^{3} \delta_{pk} \, . \nn 
\end{align}
Since the system is confined in a box, the only singularities of the finite-volume amplitudes are poles on the real energy axis, known as L\"uscher poles, which correspond to the spectrum of the system. The quantization condition allows us to connect the finite-volume spectrum of three particles in a finite spatial volume to the infinite volume scattering quantities governed by the on-shell relations. 

Since Lorentz invariance is broken, the resulting finite-volume amplitude will depend not only on the invariant mass of the two particles, but also the direction of the total momentum of the system. Total angular momentum is no longer conserved in a finite-volume, and therefore is it not a suitable quantum number to label for the two particle state, requiring the need to subduce the system to an irreducible representation of the octahedral group or one of its little groups for moving frames. We will not discuss the subduction to the cubic irreps, and instead refer to Refs.~\cite{Thomas:2011rh,Moore:2005dw} for the discussion.

The finite-volume scaling for short-distance objects will be such that they are equivalent to the infinite-volume quantity up to some correction which is exponenitally suppressed in the volume. For example, the finite-volume hadron mass $M_L$, $M_L = M + \Oc(e^{-mL})$ where $m$ is the lightest mass scale of the system, \eg the pion in QCD~\cite{Luscher:1985dn}. Similar arguments hold for all connected $K$ matrices. Therefore, we can ignore the volume correction for large enough box sizes and simply replace all short-distance finite-volume objects with their infinite-volume counterparts. This argument holds because any such object can be represented by an analytic function in the kinematic region of study since all the relevant on-shell physics has been isolated~\cite{Kim:2005gf}. Since the functions are analytic, any integral representation will not contain singular behavior in the integrand $f$, and we can freely replace discrete sums with continuous integrals up to the exponentially suppressed behavior according to the Poisson summation formula,
\begin{align}
	\frac{1}{L^{3}}\sum_{\k} \, f(\k) = \int \frac{\diff^3\k}{(2\pi)^{3}} \, f(\k) + \Oc(e^{-mL}) \, . \nn
\end{align}
%

\subsection{Two particle quantization condition}
\label{sec:QC2}

First we establish the quantization condition for the two-particle subsystem. Let us fix the spectator momentum $\k$, and work with the $\2\to\2$ amplitude $\Mc_2(\sigma_k)$. Since rotational invariance is broken, the finite-volume counterpart amplitude, denoted $\Mc_{2,L}$, will in general be a function of the total momentum flow $P_k$. Additionally the amplitude is a dense matrix in angular momentum space since angular momentum is no longer conserved. We write the finite-volume equivalent of Eq.~\eqref{eq:M2_onshell} as
\begin{align}
	\label{eq:M2L}
	\Mc_{2,L}(P_k) = \Kc_{2}^{\,\prime}(\sigma_k) - \Kc_{2}^{\,\prime}(\sigma_k) \, \Ic_{L}^{\,\prime}(P_k) \, \Mc_{2,L}(P_k) \, ,
\end{align}
where we have used the relation $\Kc_{2,L}^{\,\prime}(P_k) = \Kc_2^{\,\prime}(\sigma_k)$ which is valid in the elastic two-particle scattering region and dropping the exponentially suppressed correction with a large enough box size. Since $\Mc_{2,L}$ must only contain real-axis pole singularities in the complex energy plane, the explicit cut structure of Eq.~\eqref{eq:Ifcn} makes it difficult to transition to a finite $L$ system. We have chosen the scheme where we use the finite-volume counterpart of Eq.~\eqref{eq:Iprimefcn} since the dispersive version is more amenable to the appropriate analytic structure of the finite-volume amplitude. Note that we can rewrite Eq.~\eqref{eq:Iprimefcn} as
\begin{align}
	\label{eq:Ifcn_alt}
	\Ic_{\ell' m_{\ell' } , \ell m_{\ell}}^{\,\prime}(\sigma_k) & = \frac{ \xi }{ 2 }  \int^{\Lambda}\! \frac{\diff^3\a}{(2\pi)^{3}} \left( \frac{a^{\star}}{q_k^{\star}}  \right)^{\ell'} \frac{4\pi \, Y_{\ell' m_{\ell'}}^{*}(\bh{\a}^{\star}) Y_{\ell m_{\ell}}(\bh{\a}^{\star})}{2\omega_{a}( q_{k}^{\star\,2} - a^{\star\,2} + i\epsilon ) } \left( \frac{a^{\star}}{q_k^{\star}}  \right)^{\ell} \, ,
\end{align}
which can be seen by integrating the above expression. The re-introduction of the spherical harmonic is necessary as we will now replace the integration with a discrete sum, breaking rotational invariance. The finite-volume analogue of Eq.~\eqref{eq:Ifcn_alt} is then
\begin{align}
\label{eq:ILfcn}
\Ic_{L \, ; \, \ell' m_{\ell' } , \ell m_{\ell}}^{\,\prime}(P_k) & =\frac{ \xi }{ 2 }  \frac{1}{L^{3}} \sum_{\a}^{\Lambda} \left( \frac{a^{\star}}{q_k^{\star}}  \right)^{\ell'}  \frac{4\pi \, Y_{\ell' m_{\ell'}}^{*}(\bh{\a}^{\star}) Y_{\ell m_{\ell}}(\bh{\a}^{\star})}{2\omega_a( q_{k}^{\star\,2} - a^{\star\,2}  ) } \left( \frac{a^{\star}}{q_k^{\star}}  \right)^{\ell} \, ,
\end{align}

We can easily solve the matrix relation Eq.~\eqref{eq:M2L} for $\Mc_{2,L}$, and identify that the pole singularities occur when
\begin{align}
\label{eq:QC2}
\det \Big[ \, 1 + \Kc_2^{\,\prime}(\sigma_k) \, \Ic_L^{\,\prime}(P_k)  \,\Big]_{E_k = E_{k,\mathfrak{n}} }= 0 \, ,
\end{align}
where the determinant is over angular momentum space. Equation~\eqref{eq:QC2} is the L\"uscher quantization condition, which holds at the finite volume energies $E_{k,\mathfrak{n}}$. It is a determinant is over all angular momentum space, and provides a mapping between the infinite-volume $K$ matrices and the finite volume spectra. Formally, Eq.~\eqref{eq:QC2} holds in the kinematic region that the scattering equation was constructed, \ie for $2m \le \sqrt{\sigma_k} < 4m^2$. While Eq.~\eqref{eq:QC2} is an adequate expression, often one wants to find a regulator independent relation that is independent on the cutoff $\Lambda$. We can relate $\Kc_2^{\,\prime}$ to $\Kc_2$ via $\Kc_{2}^{\prime\,-1} = \Kc_{2}^{-1} - \re\Ic^{\,\prime}$, where $\Kc_2$ is defined with the usual two-body phase space function $\Ic$, Eq.~\eqref{eq:Ifcn}. Substituting this relation in Eq.~\eqref{eq:M2L}, we find an alternative relation
\begin{align}
	\label{eq:M2L_alt}
	\Mc_{2,L}(P_k) = \frac{1}{1 + \Kc_{2} (\sigma_k) \, F_{2,L}(P_k)} \, \Kc_{2}(\sigma_k) \, , 
\end{align}
which has a quantization condition
\begin{align}
	\det \Big[ \, 1 + \Kc_2(\sigma_k) \, F_{2,L}(P_k)  \,\Big]_{E_k = E_{k,\mathfrak{n}} }= 0 \, . \nn 
\end{align}
Here we have defined the finite-volume function
\begin{align}
	\label{eq:Ffcn}
	F_{2,L \, ; \, \ell' m_{\ell'} , \ell m_{\ell}}(P_k) & \equiv \Ic_{L \, ; \, \ell' m_{\ell'} , \ell m_{\ell}}^{\,\prime}(P_k) - \re\Ic_{\ell' m_{\ell'} , \ell m_{\ell}}^{\,\prime}(\sigma_k) \, \nn \\[5pt]
	& = \frac{ \xi }{ 2 } \left[\, \frac{1}{L^{3}} \sum_{\a}^{\Lambda} - \Pc\int^{\Lambda} \! \frac{\diff^3\a}{(2\pi)^3} \, \right] \,\left( \frac{a^{\star}}{q_k^{\star}}  \right)^{\ell'}  \frac{4\pi \, Y_{\ell' m_{\ell'}}^{*}(\bh{\a}^{\star}) Y_{\ell m_{\ell}}(\bh{\a}^{\star})}{2\omega_a^{\star}( q_{k}^{\star\,2} - a^{\star\,2}  ) } \left( \frac{a^{\star}}{q_k^{\star}}  \right)^{\ell} \, ,
\end{align}
where $\Pc$ indicates the principle value prescription.~\footnote{Note that formally $F_{2,L}$ is independent of the cutoff $\Lambda$ as $\Lambda \to \infty$, however in practice one usually requires some choice to numerically compute the object.} The function $F_{2,L}$ is equivalent to the $F_{\pv}$ functions of Refs.~\cite{Hansen:2014eka}, which can be proven using the Poisson summation formula to generate equalities up to the ignored exponentially suppressed corrections~\cite{Kim:2005gf}. We choose to use $F$ instead of $F_{\pv}$ in order to simplify the notation for this work. In effect, defining $F_{2,L}$ in this way creates a rule for replacing the two-body phase space by a finite-volume counterpart
\begin{align}
	\Ic_{\ell' m_{\ell'} , \ell m_{\ell}}(\sigma_k) = -i \delta_{\ell'\ell} \delta_{m_{\ell'} m_{\ell}} \, \frac{\xi q_k^{\star}}{8\pi\sqrt{\sigma_k}} \longrightarrow F_{2,L \, ; \, \ell' m_{\ell'} , \ell m_{\ell}}(P_k) \, . \nn 
\end{align}
%

\subsection{Three-particle quantization condition}
\label{sec:QC3}

We now construct the three-particle quantization condition. The most direct way is to use the on-shell equation for the $T$ matrix element, Eq.~\eqref{eq:Tuu_onshell}. In transitioning to the finite-volume, the finite-volume amplitudes in general will become dense matrices in both angular momentum ($\ell, m_{\ell}$) and the spectator momentum spaces ($k$). Thus every object will be cast into the $(k \ell m_{\ell})$ basis. The spectator delta function $\delta(\p,\k)$ must be replaced as
\begin{align}
	\delta(\p,\k) \to 2\omega_k L^3 \, \delta_{pk}= [\, 2\omega L^3 \, ]_{pk} \, . \nn 
\end{align}
The $[2\omega L^3]$ matrix is also trivially proportional to the identity matrix in $(\ell,m_{\ell})$ space. Integration measures over the spectator momentum then transition as
\begin{align}
	\int_k = \int\!\frac{\diff^3\k}{(2\pi)^3\,2\omega_k} \longrightarrow \sum_{\k} \, \frac{1}{2\omega_k L^3} \, . \nn
\end{align}

We denote the finite-volume $T$ matrix element as $\Tc_{L}^{(u,u)}(P)$, which is a matrix in $(k\ell m_{\ell})$ space, and a function of the total energy-momentum $P = (E,\P)$. From Eq.~\eqref{eq:Tuu_onshell}, we find that $\Tc_{L}^{(u,u)}(P)$ satisfies the relation
\begin{align}
	\Tc_{L}^{(u,u)}(P) = \Kc_{3,L}^{(u,u)}(P) - \Kc_{3,L}^{(u,u)}(P) \cdot \frac{1}{2\omega L^3} \cdot \Gamma_L(P) \cdot \frac{1}{2\omega L^3} \cdot \Tc_{L}^{(u,u)}(P)
\end{align}
where we have introduced the $\cdot$ notation to indicate the matrix product over the combined ($k\ell m_{\ell}$) space. We also use the notation that any object with an $L$ subscript depends explicitly on the box size. Implicitly all three-body objects contain an $L$ dependence through the quantized momentum, $\k = 2\pi \, \n / L$. The three-body $K$ matrix has an explicit $L$ dependence since it contains a spectator delta function,
\begin{align}
	\Kc_{3,L}^{(u,u)}(P) = 2\omega L^3 \cdot \Kc_2 + \Kc_{3,\df}^{(u,u)}(P) \, , \nn 
\end{align}
where $\Kc_2$ is the infinite-volume two-body $K$ matrix interpreted now as a diagonal matrix in ($k\ell m_{\ell}$) space where the $k$ index is encoded in the argument $\sigma_k = (P-k)^2$,  that is
\begin{align}
	[\, \Kc_2 \, ]_{p\ell'm_{\ell'},k\ell m_{\ell}} = \delta_{pk} \delta_{\ell'\ell} \delta_{m_{\ell'}m_{\ell}}\, \Kc_{2,\ell}(\sigma_k) \, , \nn 
\end{align}
 and the infinite-volume connected three-body $K$ matrix $\Kc_{3,\df}^{(u,u)}(P)$ is a dense matrix in ($k\ell m_{\ell}$) space where the elements in momentum space are determined by $\Kc_{3,\df}^{(u,u)}(\p,\k)$ with $\p,\k = 2\pi \, \n / L$. We keep $\Kc_2$ an implicit function of $P$ for convenience. Following Ref.~\cite{Blanton:2020gha} we define a convenient matrix
\begin{align}
	\wt{\Kc}_{2,L} \equiv 2\omega L^3 \cdot \Kc_2 \nn 
\end{align}
because the combination appears frequently.

The remaining finite-volume function $\Gamma_L$ is defined as
\begin{align}
	\Gamma_L(P) =  2\omega L^3 \cdot H \cdot F_{2,L} + G_L(P) \, ,
\end{align}
which is found by replacing $\Ic$ with $F_{2,L}$ and taking $H$ as a diagonal matrix in $k$ space with elements $H(\sigma_k)$. The second function is the finite-volume counterpart to $\Gc$, which is identical to the infinite-volume since $\Gc$ contains only simple poles which carry over to the finite-volume, \ie $G_L = \Gc$. The matrix elements of $G_L$ in momentum space are given by $\Gc(\p,\k)$ where $\p,\k$ are evaluated for quantized momenta. Like $\wt{\Kc}_{2,L}$, these function frequently appear with multiplicative factors of $[2\omega L^3]^{-1}$. We therefore conveniently define the functions
\begin{align}
	\wt{\Gamma}_L(P) & \equiv \frac{1}{2\omega L^3} \cdot \Gamma_L(P) \cdot \frac{1}{2\omega L^3} \, , \nn \\[5pt]
	& = \wt{F}_{2,L} + \wt{G}_L(P) \, , \nn
\end{align}
where we have defined $\wt{F}_{2,L}$ by 
\begin{align}
	\wt{F}_{2,L} \equiv \frac{1}{2\omega L^3} \cdot H\cdot F_{2,L} \, , \nn 
\end{align}
and $\wt{G}_L$ as
\begin{align}
	\wt{G}_L(P) \equiv \frac{1}{2\omega L^3} \cdot G_L(P) \cdot \frac{1}{2\omega L^3} \, . \nn
\end{align}

The L\"uscher poles of $\Tc_L^{(u,u)}$ occur whenever
\begin{align}
	\label{eq:QC3}
	\det \Big[ \, 1 + \Kc_{3,L}^{(u,u)}(P) \cdot \wt{\Gamma}_L(P)  \,\Big]_{E = E_{\mathfrak{n}} }= 0 \, ,
\end{align}
or equivalently the result shown in Eq.~\eqref{eq:QC3_v1}, which we repeat here
\begin{align}
	\det \Big[ \, 1 + \left( \wt{\Kc}_{2,L} + \Kc_{3,\df}^{(u,u)}(P)  \right) \cdot \left( \wt{F}_{2,L} + \wt{G}_L(P)\right ) \,\Big]_{E = E_{\mathfrak{n}} }= 0 \, . \nn 
\end{align}
The quantization condition is a determinant over $(k\ell m_{\ell})$ space, and agrees with the TOPT result of Ref.~\cite{Blanton:2020gha}. If one can determine the spectrum $E_{\mathfrak{n}}$ of three particles in a box, say from lattice QCD simulations, then Eq.~\eqref{eq:QC3} can be used to constrain the connected $\3\to\3$ $K$ matrix given the $\2\to\2$ $K$ matrix (or constrain both objects simultaneously). Note that if instead of Eq.~\eqref{eq:Tuu_onshell}, we use \eqref{eq:M3_onshell} to construct the finite-volume amplitude, we would arrive at the same quantization condition as \eqref{eq:QC3} since any other pole in the expression corresponds to the free three-particle spectrum.

The $K$ matrices can parameterized by some analytic function which are fitted to the spectra, and then in turn can be fed into the scattering equations such as Eq.~\eqref{eq:Tuu_onshell} or Eq.~\eqref{eq:M3_onshell} to determine the infinite-volume $\3\to\3$ scattering amplitude. Both the two and three particle quantization conditions involve infinite-dimensional matrices in angular momentum space, while the cutoff functions truncate the spectator momentum space for the three body conditions. the barrier suppression of partial wave amplitudes allow us to truncate the determinant by focusing on a low-energy region and neglecting waves which have negligible. In each case this leads to systematic effects in determining $\Kc_2$ and $\Kc_{3,\df}^{(u,u)}$, but in the $\3\to\3$ case there exists additional systematic effects since unitarity is explicitly broken since Eq.~\eqref{eq:Tuu_unitarity} and Eq.~\eqref{eq:M3_unitarity} are not diagonal in $(\ell m_{\ell})$ space. This gives another set of systematic errors which are similar to the ones usually found in isobar truncations used in phenomenological analyses of three-body decays~\cite{JPAC:2019ufm,Albaladejo:2019huw,JPAC:2021rxu}. A possible course of action is to include enough waves to saturate the system in some low-energy domain with the relevant physical channels, and treat the breaking of unitarity as a systematic error in the determination of the $\3\to\3$ amplitudes.

This concludes the main results, where we have used $S$ matrix unitarity to construct both the infinite-volume scattering equations in terms of on-shell quantities and some unknown short-distance object, and the finite-volume quantization conditions which relate these short-distance objects to the two and three particle spectra. In the following section, we present an alternative derivation which is closely related to the approach taken by Ref.~\cite{Mikhasenko:2019vhk}, and connect the symmetric $\Kc_{3,\df}$ form of Ref.~\cite{Hansen:2015zga} to the one presented here.

\section{Alternative representations}
\label{sec:alternative}

\begin{figure}[t!]
    \centering
    \includegraphics[ width=0.95\textwidth]{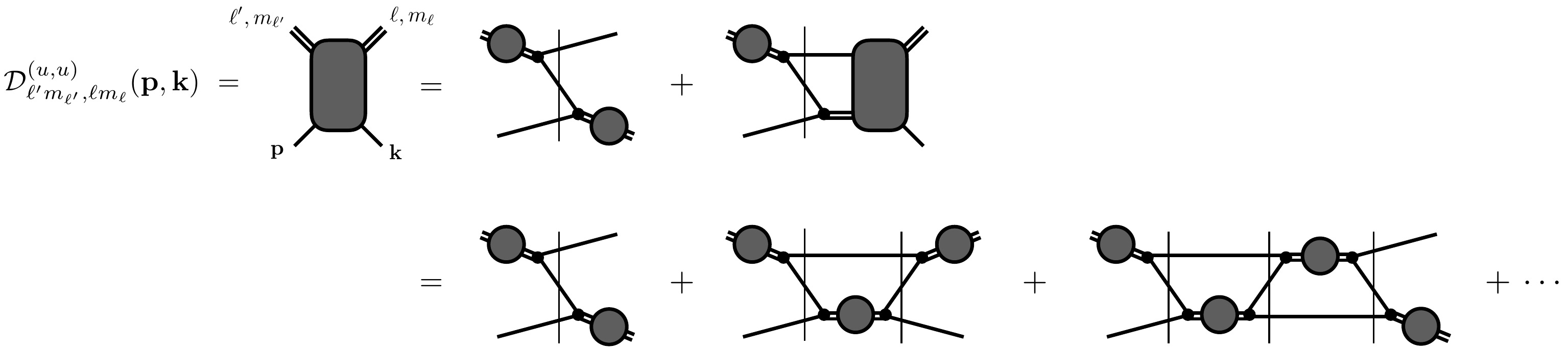}
    \caption{Exchange dominated, or ladder $\3\to\3$ amplitude $\Dc^{(u,u)}$ as defined in Eq.~\eqref{eq:ladder}. The solid vertical lines indicate that the intermediate states are represented by on-shell kinematic functions.}
    \label{fig:ladder}
\end{figure}

In this section, we derive an alternative representation of the $\3\to\3$ scattering amplitude and the subsequent quantization condition. The idea is to separate physics that is dominated by on-shell particle exchanges between two-body sub-channels from interactions which involve short-distance three-body physics. By doing so, we find the resulting framework to be similar to the one originally presented in Ref.~\cite{Hansen:2014eka,Hansen:2015zga}, with the exception that we work with the asymmetric $\Kc_{3,\df}^{(u,u)}$ instead of the symmetric form called $\Kc_{3,\df}$. Our starting point is Eq.~\eqref{eq:M3_onshell}, the on-shell representation for the connected $\3\to\3$ amplitude, which is repeated here for convenience
\begin{align}
	\Mc_{3}^{(u,u)}(\p,\k) & = \Kc_{3,\df}^{(u,u)}(\p,\k) - \Kc_{2}(\sigma_{p}) \, \Gc(\p,\k) \, \Mc_{2}(\sigma_{k}) \nn \\[5pt]
	& - \int_{k'} \Kc_{3,\df}^{(u,u)}(\p,\k') \, \Gamma(\k',\k) \, \Mc_{2}(\sigma_{k}) - \Kc_{2}(\sigma_{p}) \int_{k'} \Gamma(\p,\k') \, \Mc_{3}^{(u,u)}(\k',\k)   \nn \\[5pt]
	& - \int_{p'}\int_{k'} \Kc_{3,\df}^{(u,u)}(\p,\p') \, \Gamma(\p',\k') \, \Mc_{3}^{(u,u)}(\k',\k) \, . \nn 
\end{align}
We then multiply relation on the left by $1 - \Mc_2(\sigma_p) \, \Ic(\sigma_p)$ and use Eq.~\eqref{eq:M2_onshell} to convert the two occurrences of $\Kc_2$ to $\Mc_2$, finding
\begin{align}
	\label{eq:tmp}
	\Mc_{3}^{(u,u)}(\p,\k) & = -\Mc_2(\sigma_p) \, \Gc(\p,\k) \, \Mc_2(\sigma_k) - \Mc_2(\sigma_p) \int_{k'} \Gc(\p,\k') \, \Mc_{3}^{(u,u)}(\k',\k) \nn \\
	& + \left[ \, 1 - \Mc_2(\sigma_p) \, \Ic(\sigma_p) \, \right] \int_{k'} \Kc_{3,\df}^{(u,u)}(\p,\k') \left[ \, \delta(\k',\k) - \Gamma(\k',\k) \Mc_{2}(\sigma_k)  \, \right] \nn \\
	& - \left[ \, 1 - \Mc_2(\sigma_p) \, \Ic(\sigma_p) \, \right] \int_{p'}\int_{k'} \Kc_{3,\df}^{(u,u)}(\p,\p') \, \Gamma(\p',\k') \, \Mc_{3}^{(u,u)}(\k',\k) \, . 
\end{align}
We now introduce an amplitude $\Dc^{(u,u)}$, which is the solution for $\Mc_{3}^{(u,u)}$ when no short-distance physics is included, \ie when $\Kc_{3,\df}^{(u,u)} = 0$ or equivalently
\begin{align}
	\lim_{\Kc_{3,\df} \to 0} \Mc_{3}^{(u,u)}(\p,\k) \equiv \Dc^{(u,u)}(\p,\k) \, . \nn
\end{align}
This of course is scheme dependent statement, as any redefinition of the $K$ matrices will result in a different interpretation of the short-distance physics. However, here we fix to a specific scheme, and define $\Dc^{(u,u)}$ with respect to this scheme. The $\Dc^{(u,u)}$ amplitude is often called the \emph{ladder} amplitude, as it represents an infinite series of exchanges between $\2\to\2$ subprocesses. Therefore, when $\Kc_{3,\df}^{(u,u)} = 0$, we say there is no short-distance three-body interactions, and the system is dominated by on-shell exchanges between pair-wise interactions. Setting $\Kc_{3,\df}^{(u,u)} = 0$ in Eq.~\eqref{eq:tmp}, we find that $\Dc^{(u,u)}$ is the solution to the integral equation
\begin{align}
	\label{eq:ladder}
	\Dc^{(u,u)}(\p,\k) = -\Mc_2(\sigma_p) \, \Gc(\p,\k) \, \Mc_2(\sigma_k) - \Mc_2(\sigma_p) \int_{k'} \Gc(\p,\k') \, \Dc^{(u,u)}(\k',\k) \, ,
\end{align}
which is completely controlled by two-body physics and is represented in Fig.~\ref{fig:ladder}.
%

\begin{figure}[t!]
    \centering
    \includegraphics[ width=0.95\textwidth]{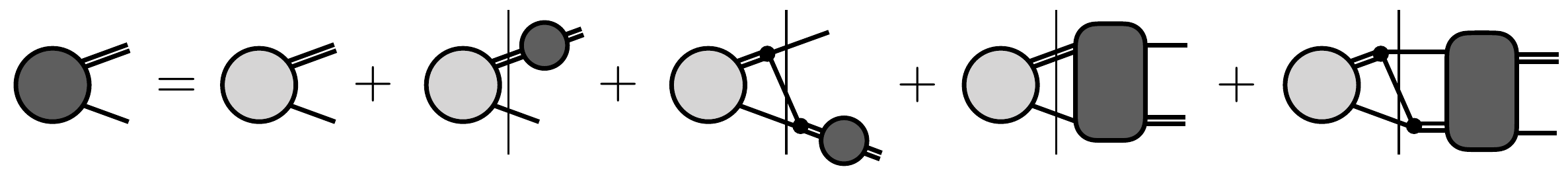}
    \caption{Diagrammatic representation of the rescattering function $\Rc^{(u,u)}$ acting on the amputated amplitude $\Tc_{\df}^{(u,u)}$ (gray filled circle) to form the divergence free amplitude $\Mc_{3,\df}^{(u,u)}$ (black filled circle).}
    \label{fig:rescattering}
\end{figure}

We can remove the ladder contribution from the integral equation for $\Mc_3^{(u,u)}$ by defining the divergence free amplitude $\Mc_{3,\df}^{(u,u)}$
\begin{align}
	\label{eq:M3df}
	\Mc_{3,\df}^{(u,u)}(\p,\k) \equiv \Mc_{3}^{(u,u)}(\p,\k) - \Dc^{(u,u)}(\p,\k) \, . 
\end{align}
Inserting this into Eq.~\eqref{eq:tmp} and using the ladder equation \eqref{eq:ladder}, we find an integral equation for $\Mc_{3,\df}^{(u,u)}$,
\begin{align}
	\Mc_{3,\df}^{(u,u)}(\p,\k) & = -\Mc_{2}(\sigma_p) \int_{k'} \Gc(\p,\k') \, \Mc_{3,\df}^{(u,u)}(\k',\k) \nn \\
	& + \left[ \, 1 - \Mc_2(\sigma_p) \, \Ic(\sigma_p)  \, \right] \int_{k'} \Kc_{3,\df}^{(u,u)}(\p,\k') \, \Rc^{(u,u)}(\k',\k) \nn \\
	& - \left[ \, 1 - \Mc_2(\sigma_p) \, \Ic(\sigma_p)  \, \right] \int_{p'}\int_{k'} \Kc_{3,\df}^{(u,u)}(\p,\p') \, \Gamma(\p',\k') \, \Mc_{3,\df}^{(u,u)}(\k',\k) \, , \nn 
\end{align}
where $\Rc^{(u,u)}$ defined as
\begin{align}
	\label{eq:Rfcn}
	\Rc^{(u,u)}(\p,\k) & = \delta(\p,\k) - \int_{k'} \Gamma(\p,\k') \, \Big[ \, \delta(\k',\k) \, \Mc_{2}(\sigma_{k'}) + \Dc^{(u,u)}(\k',\k) \, \Big] \, .
\end{align}
We interpret $\Rc^{(u,u)}$ as an initial state rescattering function, which contains three distinct topological structures of three interacting particles in the initial state before they succumb to any short-distance three-body forces. The first term of Eq.~\eqref{eq:Rfcn} represents the case where the initial state does not interact, the second indicating two of the particles scatter either directly or with an initial exchange, and the third with pair-wise exchanges. Figure~\ref{fig:rescattering} shows the different on-shell topologies which dress the initial state.

We can define a similar rescattering function for the final state, denoted by $\Lc^{(u,u)}$, which we define as
\begin{align}
	\label{eq:Lfcn}
	\Lc^{(u,u)}(\p,\k) & =  \delta(\p,\k) - \int_{k'}  \Big[ \, \delta(\p,\k') \, \Mc_{2}(\sigma_{k'}) + \Dc^{(u,u)}(\p,\k') \, \Big] \, \Gamma(\k',\k) \, ,
\end{align}
and has a similar interpretation as the initial state rescattering function. The functions $\Rc^{(u,u)}$ and $\Lc^{(u,u)}$ are closely related to those presented in Ref.~\cite{Hansen:2015zga}, which we will show in the following section.

The introduction of the initial and final state rescattering functions motivates us to define an amputated divergence-free amplitude $\Tc_{\df}^{(u,u)}$, which is free of rescatterings in both the initial and final state,
\begin{align}
	\label{eq:amputate}
	\Mc_{3,\df}^{(u,u)}(\p,\k) \equiv \int_{p'}\int_{k'} \, \Lc^{(u,u)}(\p,\p')\, \Tc_{\df}^{(u,u)}(\p',\k')\,\Rc^{(u,u)}(\k',\k)\, ,
\end{align}
Substituting this relation into the above integral equation, we find that
\begin{align}
	\int_{p'}\int_{k'} & \left[ \, 1 + \Mc_{2}(\sigma_p) \, \Gc(\p,\p') \, \right] \, \Lc^{(u,u)}(\p',\k') \, \Tc_{\df}^{(u,u)}(\k',\k) \nn \\
	& = \left[ \, 1 - \Mc_2(\sigma_p) \, \Ic(\sigma_p)  \, \right] \Kc_{3,\df}^{(u,u)}(\p,\k) \nn \\
	& -  \left[ \, 1 - \Mc_2(\sigma_p) \, \Ic(\sigma_p)  \, \right] \int_{p'}\int_{q'} \int_{k'} \Kc_{3,\df}^{(u,u)}(\p,\p') \, \Gamma(\p',\q') \, \Lc^{(u,u)}(\q',\k') \, \Tc_{\df}^{(u,u)}(\k',\k) \, . \nn
\end{align}
Combining the equations of $\Lc^{(u,u)}$, $\Dc^{(u,u)}$, and $\Mc_2$, we find the identity
\begin{align}
	\int_{k'} \left[ \, 1 + \Mc_{2}(\sigma_p) \, \Gc(\p,\k') \, \right] \, \Lc^{(u,u)}(\k',\k) & = \left[ \, 1 - \Mc_2(\sigma_p) \, \Ic(\sigma_p)  \, \right] \delta(\p,\k) \, . \nn 
\end{align}
Therefore, all terms of the above integral equation have a common $[1 - \Mc_2(\sigma_p) \, \Ic(\sigma_p) ]$ factor on the final state. Removing this factor, we arrive at the integral equation for $\Tc_{\df}^{(u,u)}$
\begin{align}
	\label{eq:Tdf_onshell}
	\Tc_{\df}^{(u,u)}(\p,\k) = \Kc_{3,\df}^{(u,u)}(\p,\k) - \int_{p'} \int_{q'} \int_{k'} \, \Kc_{3,\df}^{(u,u)}(\p,\p') \, \Gamma(\p',\q') \, \Lc^{(u,u)}(\q',\k') \, \Tc_{\df}^{(u,u)}(\k',\k) \, .
\end{align}
The structure of Eq.~\eqref{eq:Tdf_onshell} similar to the original one for the $T$ matrix element Eq.~\eqref{eq:Tuu_onshell}, except explicit kinematic divergences are removed. This includes initial and final state rescatterings, which also removes the disconnected structure of $\Mc_2$ implicitly since it is part of the rescattering functions. Equation~\eqref{eq:Tdf_onshell} also shows explicit intermediate state interactions with $\Gamma \cdot \Lc^{(u,u)}$. We can interpret $\Tc_{\df}$ as the $T$ matrix element where the initial and final state rescatterings associated with two-body systems have been removed. We have decomposed the single integral equation Eq.~\eqref{eq:M3_onshell} into two separate equations ~\eqref{eq:ladder} and \eqref{eq:Tdf_onshell}. Thus we can first study a system in the case where there are no short-distance three-body interactions with solutions of Eq.~\eqref{eq:ladder}, and then feed that into \eqref{eq:Tdf_onshell} by introducing some $\Kc_{3,\df}^{(u,u)}$. It is not yet known if working with the two separate equations or directly with \eqref{eq:M3_onshell} directly is simpler, however it presents an opportunity to study the stability of such solutions or examine different physical cases, \ie a small $\Kc_{3,\df}$ which is independent of the spectator momenta leads to an analytic solution for $\Tc_{\df}$, \cf the isotropic approximation of Ref.~\cite{Hansen:2014eka}.

%

In a finite-volume, we can write the counterpart to $\Tc_{\df}^{(u,u)}$ as
\begin{align}
	\Tc_{\df,L}^{(u,u)}(P) = \Kc_{3,\df}^{(u,u)}(P) -  \Kc_{3,\df}^{(u,u)}(P) \cdot \wt{\Gamma}_L(P) \cdot \wt{\Lc}_{L}^{(u,u)}(P) \cdot \Tc_{\df,L}^{(u,u)}(P) \, ,  
\end{align}
where we have defined a new finite-volume function
\begin{align}
	\wt{\Lc}_{L}^{(u,u)}(P) & \equiv \Lc_{L}^{(u,u)}(P) \cdot \frac{1}{2\omega L^3} \, , \nn \\[5pt]
	& \equiv \left[ 2 \omega L^3 -  \left( 2\omega L^3 \cdot \Mc_{2,L} + \Dc_{L}^{(u,u)} (P) \right) \cdot \frac{1}{2\omega L^3} \cdot \Gamma_L(P) \right] \cdot \frac{1}{2\omega L^3} \, , \nn \\[5pt]
	& = 1 - 2\omega L^3 \cdot \Mc_{2,L} \cdot \wt{\Gamma}_L(P) - \Dc_L^{(u,u)}(P) \cdot \wt{\Gamma}_L(P) \, , \nn 
\end{align}
where $\Mc_{2,L}$ is given by Eq.~\eqref{eq:M2L_alt} and $\Dc_{L}^{(u,u)}$ is defined by
\begin{align}
	\Dc_L^{(u,u)}(P) & = -\Mc_{2,L} \cdot G_L(P) \cdot \Mc_{2,L} - \Mc_{2,L} \cdot G_L(P) \cdot \frac{1}{2\omega L^3} \cdot \Dc_{L}^{(u,u)}(P) \, ,
\end{align}
which follows from Eq.~\eqref{eq:ladder} as its finite-volume counterpart. One can show by direct manipulation that the combination $2\omega L^3 \cdot \Mc_{2,L} + \Dc_{L}^{(u,u)} (P)$ can be simplified to
\begin{align}
	2\omega L^3 \cdot \Mc_{2,L} + \Dc_{L}^{(u,u)} (P) = \frac{1}{{\wt{\Kc}_{2,L}}^{\, -1} + \wt{F}_{2,L} + \wt{G}_{L}(P)} \, , \nn 
\end{align}
thus the combination $\Gamma_L \cdot \wt{\Lc}^{(u,u)}_L$ can be written entirely in terms of $\wt{\Kc}_{2,L}$, $\wt{F}_{2,L}$, and $\wt{G}_L$. We define a convenient the function $F_{3,L}^{(u,u)}$ as
\begin{align}
	F_{3,L}^{(u,u)}(P) & \equiv \wt{\Gamma}_L(P) \cdot \wt{\Lc}_L^{(u,u)}(P) \, , \nn \\[5pt]
	& = \left[\,\wt{F}_{2,L} + \wt{G}_{L}(P)\,\right] \left( \, 1 - \frac{1}{{\wt{\Kc}_{2,L}}^{\, -1} + \wt{F}_{2,L} + \wt{G}_{L}(P)} \cdot \left[\,\wt{F}_{2,L} + \wt{G}_{L}(P)\,\right] \right) \, ,
\end{align}
which was given in Eq.~\eqref{eq:F3uu}.

The L\"uscher poles occur when the inverse of $\Mc_{3}^{(u,u)}$ vanishes. Since the rescattering functions do not contain the three-body $K$ matrix, the spectrum for three interacting particles can equally be found when the inverse of $\Tc_{\df,L}^{(u,u)}$ vanishes. Thus, an alternative yet equivalent form for the quantization condition is given by Eq.~\eqref{eq:QC3_new}, for convenience repeated here
\begin{align}
	\det \Big[ \, 1 + \Kc_{3,\df}^{(u,u)}(P) \cdot F_{3,L}^{(u,u)}(P) \,\Big]_{E = E_{\mathfrak{n} } }= 0 \, .
\end{align}

In these preceding manipulations, it is presented in such a way which introduces some physical content into the various definitions amplitudes, but ultimately is just a way for us to derive a set of relations which are similar to those in other works for which we know the answer. In the next section, we shall explore how these equations arise naturally from the unitarity relations, and present and alternative derivation of Eq.~\eqref{eq:Tdf_onshell}. At the end of the alternative derivation, we make a connection to the form of the original FVU result~\cite{Mai:2017vot,Mai:2017bge}, and show how the difference manifest by considering different rescattering functions. An advantage of this new form is it allows us to convert to the original results of Refs.~\cite{Hansen:2014eka,Hansen:2015zga}, which use a symmetric form of the $K$ matrix called $\Kc_{3,\df}$ instead of working with the asymmetric $\Kc_{3,\df}^{(u,u)}$. We show how to obtain the symmetric form of Eq.~\eqref{eq:Tdf_onshell} in Sec.~\ref{sec:symmetric_K}.

\subsection{Alternative derivation of the $\Tc_{\textnormal{\df}}^{(u,u)}$ integral equation}
\label{sec:alternative_derivation}

We now show how Eq.~\eqref{eq:Tdf_onshell} arises directly from the unitarity condition for $\Mc_{3}^{(u,u)}$, \ie Eq.~\eqref{eq:M3_onshell}. We schematically follow the derivation shown in Ref.~\cite{Mikhasenko:2019vhk}, which presented a variant of the $R$ matrix form from Refs.~\cite{Mai:2017vot,Jackura:2018xnx}. We will note the underlying difference in deriving the $R$ matrix form instead of the $K$ matrix form discussed here at the end of this section. The unitarity condition is repeated here for convenience,
\begin{align}
	\im \Mc_{3}^{(u,u)} (\p,\k) & =  \int_{p'} \int_{k'} \, \Mc_{3}^{(u,u)\,\dag }(\p,\p')   \, \Phi(\p',\k') \,   \Mc_{3}^{(u,u)}(\k',\k) \nn \\[5pt]
	& + \int_{k'}  \, \Mc_{2}^{\dag}(\sigma_{p}) \,  \Phi(\p,\k')  \, \Mc_{3}^{(u,u)}(\k',\k) \nn \\[5pt]
	& +  \int_{p'} \, \Mc_{3}^{(u,u)\,\dag }(\p,\p')  \, \Phi(\p',\k)  \, \Mc_{2}(\sigma_{k}) \nn \\[5pt]
	& + \Mc_{2}^{\dag} (\sigma_{p}) \,  \Delta(\p,\k) \, \Mc_{2}(\sigma_{k}) \, , \nn 
\end{align}
contains a pole singularity in the last term. Therefore, the amplitude $\Mc_{3}^{(u,u)}$ must contain this singularity explicitly~\cite{Eden:1966dnq}. This is exactly the exchange pole we introduced in Eq.~\eqref{eq:Gfcn}. Let us denote the amplitude for a single particle exchange by $\Dc_0^{(u,u)}$, and define it as
\begin{align}
	\label{eq:D0}
	\Dc_0^{(u,u)} (\p,\k) = - \Mc_{2}(\sigma_p) \, \Gc(\p,\k) \, \Mc_{2}(\sigma_k) \, ,
\end{align}
We can now remove the single exchange pole from $\Mc_{3}^{(u,u)}$, where we define
\begin{align}
	\label{eq:Mdf_0}
	\Mc_{3, 0}^{(u,u)} (\p,\k) \equiv \Mc_{3}^{(u,u)}(\p,\k) - \Dc_0^{(u,u)}(\p,\k) \, .
\end{align}
We substitute Eq.~\eqref{eq:Mdf_0} everywhere in \eqref{eq:M3_unitarity}, and use the imaginary part of \eqref{eq:D0} given by
\begin{align}
	\label{eq:ImD0}
	\im \Dc_0^{(u,u)}(\p,\k) & =  \Mc_{2}^{\dag}(\sigma_p) \, \rho(\sigma_p) \, \Dc_0^{(u,u)}(\p,\k) + \Mc_{2}^{\dag}(\sigma_p) \, \Delta(\p,\k) \, \Mc_{2}(\sigma_k) \nn \\
 & + \Dc_0^{(u,u)\,\dag}(\p,\k) \, \rho(\sigma_k) \, \Mc_{2}(\sigma_k) \, ,
\end{align}
where Eq.~\eqref{eq:M2_unitarity} was used on the first and last term. The three terms in Eq.~\eqref{eq:ImD0} are removed, yielding a new unitarity relation for $\Mc_{3,0}^{(u,u)}$
\begin{align}
	\label{eq:M30_unitarity}
	\im \Mc_{3, 0}^{(u,u)} (\p,\k) & =   \int_{p'} \int_{k'} \, \Mc_{3,0}^{(u,u)\,\dag }(\p,\p')   \, \Phi(\p',\k') \,   \Mc_{3,0}^{(u,u)}(\k',\k) \nn \\
	& +  \int_{p'} \int_{k'} \, \Mc_{3,0}^{(u,u)\,\dag }(\p,\p')   \, \Phi(\p',\k') \,   \Big[ \, \delta(\k',\k) \, \Mc_{2}(\sigma_k) +  \Dc_0^{(u,u)}(\k',\k) \, \Big] \nn \\
	& +  \int_{p'} \int_{k'} \, \Big[ \, \delta(\p,\p') \, \Mc_{2}^{\dag}(\sigma_p) + \Dc_0^{(u,u)\,\dag }(\p,\p')  \, \Big] \, \Phi(\p',\k') \,   \Mc_{3,0}^{(u,u)}(\k',\k)  \nn \\
	& + \int_{p'}\int_{k'} \, \Dc_0^{(u,u)\,\dag }(\p,\p')   \, \Phi(\p',\k') \,   \Dc_0^{(u,u)}(\k',\k) \nn \\
	&  + \int_{k'}  \, \Mc_{2}^{\dag}(\sigma_p) \,  \Delta(\p,\k')  \, \Dc_0^{(u,u)}(\k',\k) \nn \\
	& + \int_{p'} \, \Dc_0^{(u,u)\,\dag }(\p,\p')  \, \Delta(\p',\k)  \, \Mc_{2}(\sigma_k) \, .
\end{align}

The kinematic singularity associated with the single particle exchange is absent, however we have introduced explicit singularities from two particle exchanges in the last three terms. The procedure of removing isolated kinematic divergences is repeated, this time for the two particle exchange amplitude
\begin{align}
	\label{eq:D1}
	\Dc_1^{(u,u)}(\p,\k) = \Mc_{2}(\sigma_p) \, \int_{k'} \, \Gc(\p,\k') \, \Mc_{2}(k') \, \Gc(\k',\k) \, \Mc_{2}(\sigma_k) \, .
\end{align}
The imaginary part of Eq.~\eqref{eq:D1} gives five terms, three of which are contained in the last three terms of Eq.~\eqref{eq:M30_unitarity}, and two which are contained in the second and third term of Eq.~\eqref{eq:M30_unitarity}. Defining a new amplitude $\Mc_{3,1}^{(u,u)}$ which does not contain $\Dc_1^{(u,u)}$,
\begin{align}
	\Mc_{3,1}^{(u,u)}(\p,\k) & \equiv \Mc_{3,0}^{(u,u)}(\p,\k) - \Dc_{1}^{(u,u)}(\p,\k) \, , \nn \\
	& = \Mc_{3}^{(u,u)}(\p,\k) - \Dc_{0}^{(u,u)}(\p,\k)  - \Dc_{1}^{(u,u)}(\p,\k) \, , \nn 
\end{align}
and inserting into Eq.~\eqref{eq:M30_unitarity} gives a new unitarity relation now with all kinematic singularities associated with two particle exchanges removed. This will introduce new singularities arising from three particle exchanges, and the processes can be repeated \textit{ad infinitum} to generate the $n$ particle exchange amplitudes $\Dc_{n}^{(u,u)}$ which are then subsequently removed from the remaining connected $\3\to\3$ amplitude.

Removing all kinematic exchanges allows us to then extend the notion in Eq.~\eqref{eq:Mdf_0} to define the \textit{divergence free} $\3\to\3$ amplitude,
\begin{align}
	\label{eq:M3df_alt}
	\Mc_{3,\df}^{(u,u)}(\p,\k) & \equiv \Mc_{3}^{(u,u)} (\p,\k) - \sum_{n = 0}^{\infty} \Dc_{n}^{(u,u)}(\p,\k) \, ,
\end{align}
where the infinite sum of all $n$ particle exchange amplitudes, defined by the iterate
\begin{align}
	\Dc_{n}^{(u,u)}(\p,\k) = (-1)^{n} \, \Mc_{2}(\sigma_p) \, \bigg[ \,  \prod_{j = 1}^{n} \int_{k_j}  \, \bigg] \, \prod_{j=0}^{n} \bigg( \, \Gc(\k_{j},\k_{j+1}) \, \Mc_{2}(\sigma_{k_{j+1}})  \, \bigg) \Big\rvert_{\k_0 = \p, \k_{n+1} = \k } \, , \nn 
\end{align}
are removed. Equation~\eqref{eq:M3df_alt} is identical to \eqref{eq:M3df} since the infinite sum of exchanges is exactly the ladder amplitude
\begin{align}
	\Dc^{(u,u)}(\p,\k) & = \sum_{n = 0}^{\infty} \Dc_{n}^{(u,u)}(\p,\k) \, , \nn 
\end{align}
which is the solution to the integral equation Eq.~\eqref{eq:ladder}, which can be verified by successive substitutions of the equation into itself. By construction, the ladder amplitude satisfies the relation
\begin{align}
	\label{eq:Duu_unitarity}
	\im \Dc^{(u,u)} (\p,\k) & =  \int_{p'} \int_{k'} \, \Dc^{(u,u)\,\dag }(\p,\p')   \, \Phi(\p',\k') \,   \Dc^{(u,u)}(\k',\k) \nn \\
	&  + \int_{k'}  \, \Mc_{2}^{\dag}(\sigma_p) \,  \Phi(\p,\k')  \, \Dc^{(u,u)}(\k',\k) \nn \\
	& + \int_{p'} \, \Dc^{(u,u)\,\dag }(\p,\p')  \, \Phi(\p',\k)  \, \Mc_{2}(\sigma_k) \nn \\
	& + \Mc_{2} ^{\dag} (\sigma_p) \, \Delta(\p,\k) \, \Mc_{2}(\sigma_k),
\end{align}
which contains all terms removed from the $\Mc_{3}^{(u,u)}$ unitarity equation via Eq.~\eqref{eq:M3df_alt}.

All short distance three body interactions are contained within $\Mc_{3,\df}^{(u,u)}$ since the ladder amplitude represents the contributions consisting of exchanges between two particle subprocesses. After substituting Eq.~\eqref{eq:M3df_alt} into \eqref{eq:M3_unitarity} and using \eqref{eq:Duu_unitarity}, we find that $\Mc_{3,\df}^{(u,u)}$ has the unitarity relation
\begin{align}
	\label{eq:M3df_unitarity}
	\im \Mc_{3,\df}^{(u,u)} (\p,\k) & =   \int_{p'} \int_{k'} \, \Mc_{3,\df}^{(u,u)\,\dag }(\p,\p')   \, \Phi(\p',\k') \,   \Mc_{3,\df}^{(u,u)}(\k',\k) \nn \\
	& +  \int_{p'} \int_{k'} \, \Mc_{3,\df}^{(u,u)\,\dag }(\p,\p')   \, \Phi(\p',\k') \,   \Big[ \, \delta(\k',\k) \, \Mc_{2}(\sigma_k) +  \Dc^{(u,u)}(\k',\k) \, \Big] \nn \\
	& +  \int_{p'} \int_{k'} \, \Big[ \, \delta(\p,\p') \, \Mc_{2}^{\dag}(\sigma_p) + \Dc^{(u,u)\,\dag }(\p,\p')  \, \Big] \, \Phi(\p',\k') \,   \Mc_{3,\df}^{(u,u)}(\k',\k) \, .
\end{align}
The second and third terms represent singularities associated with the physics of initial and final state interactions, respectively, and are composed of entirely $\2\to\2$ sub-processes. The unitarity relation~\eqref{eq:M3df_unitarity} implies that $\Mc_{3,\df}^{(u,u)}$ must have an amputated structure, where the initial and final state of three particle can undergo pair wise interactions. The possible rescatterings are encoded in functions $\Rc^{(u,u)}$ and $\Lc^{(u,u)}$ which we defined in Eqs.~\eqref{eq:Rfcn} and \eqref{eq:Lfcn}, respectively. We remove these rescattering topologies by introducing the amputated divergence free amplitude $\Tc_{\df}^{(u,u)}$ as defined in Eq.~\eqref{eq:amputate}, and use the imaginary parts of $\Rc^{(u,u)}$ and $\Lc^{(u,u)}$ given by
\begin{align}
	\im \Lc^{(u,u)}(\p,\k) & = \int_{p'}\int_{k'} \,  \Big[ \, \delta(\p,\p') \, \Mc_{2}^{\dag}(p') + \Dc^{(u,u)\, \dag }(\p,\p') \, \Big] \,  \Phi(\p',\k') \, \Lc^{(u,u)}(\k',\k) \, ,  \nn \\[5pt]
	\im \Rc^{(u,u)}(\p,\k) & =  \int_{p'} \int_{k'}  \, \Rc^{(u,u)\,\dag}(\p,\p') \, \Phi(\p',\k') \, \Big[ \, \delta(\k',\k) \, \Mc_{2}(k) + \Dc^{(u,u)}(\k',\k) \, \Big] \, , \nn
\end{align}
to reduce the unitarity constraint to one for the $\Tc_{\df}^{(u,u)}$ amplitude,
\begin{align}
	\label{eq:Tdf_unitarity}
	\im \Tc_{\df}^{(u,u)}(\p,\k) = \int_{p'}\int_{k'} \,  \Tc_{\df}^{(u,u)\,\dag}(\p,\p') \, \Xi(\p',\k')\, \Tc_{\df}^{(u,u)}(\k',\k) \, .
\end{align}
Here we have introduced yet another new kinematic function $\Xi$, which accounts for the intermediate state singularities from all two-body rescatterings 
\begin{align}
	\Xi(\p,\k) & \equiv \int_{p'}\int_{k'} \,  \Lc^{(u,u)\,\dag}(\p,\p') \, \Phi(\p',\k') \, \Lc^{(u,u)}(\k',\k) \, , \nn \\
	& =  -\im \left[ \, \int_{p'} \,  \Gamma(\p,\p') \, \Lc^{(u,u)}(\p',\k) \, \right] \, .
\end{align}
The first line follows directly from the substitution of Eq.~\eqref{eq:amputate} into \eqref{eq:M3df_unitarity}, and is equal to the second line by direct computation of the imaginary part and using $\im \Lc^{(u,u)}$. Equation~\eqref{eq:Tdf_unitarity} is exactly the condition found in Ref.~\cite{Briceno:2019muc} when the authors showed that their original integral equations satisfied unitarity.

The unitarity equation for $\Tc_{\df}^{(u,u)}$ is similar in structure to the condition for the $\2\to\2$ amplitude, or the original $\3\to\3$ $T$ matrix unitarity relation. Therefore, an appropriate on-shell representation is given by the previously derived linear integral equation~\eqref{eq:Tdf_onshell}, which we recall as
\begin{align}
	\Tc_{\df}^{(u,u)}(\p,\k) = \Kc_{3,\df}^{(u,u)}(\p,\k) - \int_{p'} \int_{q'} \int_{k'} \, \Kc_{3,\df}^{(u,u)}(\p,\p') \, \Gamma(\p',\q') \, \Lc^{(u,u)}(\q',\k') \, \Tc_{\df}^{(u,u)}(\k',\k) \, , \nn 
\end{align}
where $\Kc_{3,\df}^{(u,u)}$ is the $\3\to\3$ connected $K$ matrix as before. If one desires, we can reintroduce the initial and final state rescattering functions to Eq.~\eqref{eq:Tdf_onshell} to yield and integral equation for $\Mc_{3,\df}^{(u,u)}$ directly,
\begin{align}
	\label{eq:M3df_onshell}
	\Mc_{3,\df}^{(u,u)}(\p,\k) & = \int_{p'}\int_{k'} \, \Lc^{(u,u)}(\p,\p')\, \Kc_{3,\df}^{(u,u)}(\p',\k')\,\Rc^{(u,u)}(\k',\k) \nn \\[5pt]
	& -  \int_{p'}\int_{q'}\int_{k'} \, \Lc^{(u,u)}(\p,\p')\, \Kc_{3,\df}^{(u,u)}(\p',\q') \, \Gamma(\q',\k') \,\Mc_{3,\df}^{(u,u)}(\k',\k) \, . 
\end{align}
It can be shown that Eq.~\eqref{eq:Tdf_onshell} satisfies Eq.~\eqref{eq:Tdf_unitarity} using the techniques in Appendix~\ref{sec:app_proof}.

\subsubsection{$B$ matrix form}

An aside, we summarize the derivation the $B$ matrix form (also called the $R$ or $C$ matrix forms in the literature) which was derived in Refs.~\cite{Mai:2017vot,Jackura:2018xnx,Mikhasenko:2019vhk}. Doing so will reveal the physical, yet inconsequential, difference between the original FVU and RFT frameowrks. Notice that Eq.~\eqref{eq:Duu_unitarity}, the unitarity relation for $\Dc^{(u,u)}$, is exactly the same as for the original connected $\3\to\3$ amplitude $\Mc_{3}$. It was argued that any chosen $\Gc$ is valid so long as $\im \Gc = -\Delta$, which means any additional arbitrary real function can be added to $\Gc$. Since it is a real function being added, the resulting alternate integral equation to Eq.~\eqref{eq:ladder} will satisfy Eq.~\eqref{eq:M3_unitarity}. This is the basis for the $R$ matrix form, where the additional unknown real function represents short-range three body physics. This new integral equation is of the same form as Eq.~\eqref{eq:ladder}, and is given by
\begin{align}
	\label{eq:Bmatrix}
	\Mc_3^{(u,u)}(\p,\k) & = \Mc_{2}(\sigma_p) \, \Bc^{(u,u)}(\p,\k) \, \Mc_{2}(\sigma_k)  + \Mc_{2}(\sigma_p) \, \int_{k'} \, \Bc^{(u,u)}(\p,\k') \, \Mc_3^{(u,u)} (\k',\k) \, ,
\end{align}
where $\Bc^{(u,u)}(\p,\k) \equiv \Cc^{(u,u)}(\p,\k) - \Gc(\p,\k)$ where $\Cc^{(u,u)}$ represents the short-distance physics, that is it plays the same role as $\Kc_{3,\df}^{(u,u)}$.~\footnote{In Ref.~\cite{Jackura:2018xnx,Mikhasenko:2019vhk,Jackura:2019bmu} the short-distance object was called $R$, while in Ref.~\cite{Mai:2017vot} it was called $C$. We choose to not use $R$ here to avoid confusion with the rescattering function.} The generalization $-\Gc(\p,\k) \longrightarrow \Bc^{(u,u)}(\p,\k)$ is allowed since we define $\Gc$ such that $\im \Gc = -\Delta$, therefore any real function ensures unitarity holds. Therefore, the relation between $\Cc$ and $\Kc_{3,\df}$ is simply a choice of zero $\Kc_{3,\df}$ and choosing a different $\Gc$ function which has an unknown real part, in which one recovers the original FVU equations from the RFT versions. It can be shown, \cf Refs.~\cite{Jackura:2019bmu,Blanton:2020jnm}, that the $\Kc_{3,\df}^{(u,u)}$ and $\Cc^{(u,u)}$ matrices are related by an integral equation,
\begin{align}
	\label{eq:Kdf_Cuu_relate}
	\Kc_{3,\df}^{(u,u)}(\p,\k) = \Kc_{2}(\sigma_p) \, \Cc^{(u,u)}(\p,\k) \, \Kc_{2}(\sigma_k) + \Kc_{2}(\sigma_p) \int_{k'} \, \Cc^{(u,u)}(\p,\k') \, \Kc_{3,\df}^{(u,u)}(\k',\k) \, , 
\end{align}
where we must use the same implicit cutoff dependence in order for the relation to hold. Consequently, this relation shows that relating these $K$ matrix objects is a non-trivial task. Initial studies using the FVU~\cite{Mai:2018djl} and the RFT~\cite{Blanton:2019vdk} forms have shown a large order of magnitude discrepancy between $\Cc$ and a related symmetric form of $\Kc_{3,\df}$, discussed in the next section. The origin of this disparity may arise from the dressings of these short-distance objects given in Eq.~\eqref{eq:Kdf_Cuu_relate}. Moreover, both analyses assume that these short-distance functions are constants in energy, which clearly violates Eq.~\eqref{eq:Kdf_Cuu_relate}. The finite-volume amplitude corresponding to Eq.~\eqref{eq:Bmatrix} is given by
\begin{align}
	\Mc_{3,L}^{(u,u)}(P) & = \Mc_{2,L} \cdot \Bc_L^{(u,u)}(P) \cdot \Mc_{2,L} + \Mc_{2,L} \cdot  \Bc_L^{(u,u)}(P) \cdot \frac{1}{2\omega L^3} \cdot \Mc_{3,L}^{(u,u)} (P) \, , \nn
\end{align}
where $\Bc_L^{(u,u)} = \Cc^{(u,u)} - G_L^{(u,u)}$ where $G_L^{(u,u)}$ is as in Sec.~\ref{sec:QC3}. The quantization condition can then be shown to be
\begin{align}
	\det \Big[ \, \wt{\Kc}_{2,L}^{\,-1} + \wt{F}_{2,L} + \wt{G}_L(P) -  \frac{1}{2\omega L^3} \cdot \Cc^{(u,u)}(P) \cdot \frac{1}{2\omega L^3} \,\Big]_{E = E_{\mathfrak{n} } }= 0 \, ,
\end{align}
which is the simple generalization of Ref.~\cite{Mai:2017bge} for arbitrary angular momenta.

We argued that any scattering equation satisfying the unitarity constraint is a valid on-shell representation, however there remains an apparent difference in the forms for the short-distance object in Eqs.~\eqref{eq:Tuu_onshell} and \eqref{eq:Bmatrix}. While both are valid representations, there is a physical difference in the initial and final state rescatterings, as discussed in Ref.~\cite{Jackura:2019bmu}. To examine the difference, we follow the approach taken in Ref.~\cite{Mikhasenko:2019vhk} which is as we presented in the preceding section until we reach Eq.~\eqref{eq:M3df_unitarity}, repeated here for convenience
\begin{align}
	\im \Mc_{3,\df}^{(u,u)} (\p,\k) & =   \int_{p'} \int_{k'} \, \Mc_{3,\df}^{(u,u)\,\dag }(\p,\p')   \, \Phi(\p',\k') \,   \Mc_{3,\df}^{(u,u)}(\k',\k) \nn \\
	& +  \int_{p'} \int_{k'} \, \Mc_{3,\df}^{(u,u)\,\dag }(\p,\p')   \, \Phi(\p',\k') \,   \Big[ \, \delta(\k',\k) \, \Mc_{2}(\sigma_k) +  \Dc^{(u,u)}(\k',\k) \, \Big] \nn \\
	& +  \int_{p'} \int_{k'} \, \Big[ \, \delta(\p,\p') \, \Mc_{2}^{\dag}(\sigma_p) + \Dc^{(u,u)\,\dag }(\p,\p')  \, \Big] \, \Phi(\p',\k') \,   \Mc_{3,\df}^{(u,u)}(\k',\k) \, . \nn 
\end{align}
The difference in the $B$ matrix form comes in at this step, where Ref.~\cite{Mikhasenko:2019vhk} defines a different rescattering function which does not contains the possibility that the particles do not interact pairwise before the short-distance three-body forces. Defining this rescattering function as $\Xc^{(u,u)}$,
\begin{align}
	\Xc^{(u,u)}(\p,\k) = \delta(\p,\k) \, \Mc_{2}(k) + \Dc^{(u,u)}(\p,\k) \, , \nn
\end{align}
and the corresponding amputated amplitude $\wt{\Tc}_{\df}^{(u,u)}$,
\begin{align}
	\Mc_{3,\df}^{(u,u)}(\p,\k) \equiv \int_{p'}\int_{k'} \, \Xc^{(u,u)}(\p,\p')\, \wt{\Tc}_{\df}^{(u,u)}(\p',\k')\,\Xc^{(u,u)} \, (\k',\k) \, . \nn
\end{align}
Since an the central difference between the rescattering functions here and the ones in Eqs.~\eqref{eq:Rfcn} and \eqref{eq:Lfcn} is an identity element,~\footnote{There also remains convolutions with kinematic factors, however careful analysis shows that this is also an inconsequential factor with respect to unitarity.} which does not contain any singularities, the difference in rescattering functions does not affect unitarity. Therefore, any affects of the initial or final state particles were absorbed by the remaining rescattering functions. We find the unitarity condition
\begin{align}
	\im \wt{\Tc}_{\df}^{(u,u)}(\p,\k) = \int_{p'}\int_{k'} \,  \wt{\Tc}_{\df}^{(u,u)\,\dag}(\p,\p') \, \im\Xc^{(u,u)}(\p',\k')\, \wt{\Tc}_{\df}^{(u,u)}(\k',\k) \, , \nn 
\end{align}
where $\im\Xc^{(u,u)}$ is given by
\begin{align}
	\im\Xc^{(u,u)}(\p,\k) = \int_{p'}\int_{k'} \Xc^{(u,u) \, \dag}(\p,\p') \, \Phi(\p',\k') \, \Xc^{(u,u)}(\k',\k) \, . \nn
\end{align}
The solution of this unitarity condition is the integral equation
\begin{align}
	\wt{\Tc}_{\df}^{(u,u)}(\p,\k) = \Cc^{(u,u)}(\p,\k) + \int_{p'}\int_{k'} \Cc^{(u,u)}(\p,\p') \, \Xc^{(u,u)}(\p',\k') \, \wt{\Tc}_{\df}^{(u,u)}(\k',\k) \, , \nn
\end{align}
which was shown to be equivalent to Eq.~\eqref{eq:Bmatrix} in Ref.~\cite{Jackura:2019bmu}. We conclude that the relation between the $\Kc_{3,\df}^{(u,u)}$ and $\Cc^{(u,u)}$ is the presence/absence of the possibility of no initial or final state rescattering, which the integral equation relating the two quantities either removes or inserts these effects depending on the desired short-distance function.

\subsection{Symmetric $K$ matrix form}
\label{sec:symmetric_K}

The final formulation we cast the scattering equations and quantization conditions to is to one which uses a symmetric $\Kc_{3,\df}$ form which was first introduced in Refs.~\cite{Hansen:2015zga} by taking the $L\to \infty$ limit of the finite volume of their results using the RFT framework. We use similar identities that were used in Ref.~\cite{Blanton:2020gha} for the finite-volume quantization conditions, but we adapt them here for the infinite-volume scattering amplitudes. To construct the symmetric $\Kc_{3,\df}$, we first introduce an alternative version of the asymmetric $\Kc_{3,\df}^{(u,u)}$ which is defined with a different exchange propagator $\Gc^{\,\prime}(\p,\k) \equiv \Gc(\p,\k) - \Gc_{\pv}(\p,\k) $ where $\Gc_{\pv}$ is the principal part of the exchange propagator. Therefore, this alternative exchange is thus
\begin{align}
	\Gc_{\ell' m_{\ell'} , \ell m_{\ell}}^{\,\prime}(\p,\k) & \equiv - i \pi \,  \delta\left( (P - k - p)^2 - m^2 \right) \, H(\sigma_p) H(\sigma_k) \, 4\pi Y_{\ell' m_{\ell'}}^{*}(\bh{\k}_{p}^{\star}) Y_{\ell m_{\ell}}(\bh{\p}_{k}^{\star}) \Big\rvert_{(P - k - p)^2 = m^2 } \, , \nn 
\end{align}
such that only the on-shell point contributes. The scattering equation for the $T$ matrix element with this choice is given by
\begin{align}
	\label{eq:Tuu_onshell_prime}
	\Tc^{(u,u)}(\p,\k) & = \Kc_{3}^{(u,u)\, \prime}(\p,\k) - \int_{p'} \int_{k'} \Kc_{3}^{(u,u)\,\prime}(\p,\p') \,  \Gamma^{\,\prime}(\p',\k')  \, \Tc^{(u,u)}(\k',\k) \, , 
\end{align}
where $\Kc_{3}^{(u,u)\, \prime}(\p,\k) = \delta(\p,\k) \, \Kc_2(\sigma_k) + \Kc_{3,\df}^{(u,u)\, \prime}(\p,\k)$ and
\begin{align}
	\Gamma^{\,\prime}(\p,\k) & = \delta(\p,\k) \, H(\sigma_k) \, \Ic(\sigma_k) + \Gc^{\,\prime}(\p,\k) \, , \nn  \\[5pt]
	& = \Gamma(\p,\k) - \Gc_{\pv}(\p,\k) \, , \nn 
\end{align}
with $\Gamma(\p,\k)$ being the original function we defined in Eq.~\eqref{eq:Gammafcn}. Since $\Tc^{(u,u)}$ in both Eqs.~\eqref{eq:Tuu_onshell} and \eqref{eq:Tuu_onshell_prime} must be the same given that they are physical amplitudes, we can relate the different $K$ matrices by equating these expressions in $\Tc^{(u,u)}$. In Appendix \ref{sec:app_K_matrices} we demonstrate the procedure, giving here the final relation
\begin{align}
	\label{eq:Kmat_relation}
	\Kc_{3,\df}^{(u,u)}(\p,\k) & = \int_{p'} \int_{k'} \Cc_L(\p,\p') \, \Kc_{3,\df}^{(u,u) \, \prime}(\p',\k') \, \Cc_R(\k',\k) \nn \\
	& + \int_{p'}\int_{q'}\int_{k'} \Cc_L(\p,\p') \, \Kc_{3,\df}^{(u,u)\, \prime}(\p',\q')  \Gc_{\pv}(\q',\k') \, \Kc_{3,\df}^{(u,u)}(\k',\k) \, ,
\end{align}
where
\begin{align}
	\Cc_L(\p,\k) & = \delta(\p,\k) + \Kc_2(\sigma_p) \, \Gc_{\pv}(\p,\k) \, , \nn \\[5pt]
	\Cc_R(\p,\k) & = \delta(\p,\k) +  \Gc_{\pv}(\p,\k) \, \Kc_2(\sigma_k) \, . \nn
\end{align}
The integral equation~\eqref{eq:Kmat_relation} is structurally identical to an analogous relationship derived in Ref.~\cite{Blanton:2020gha}. 

In order to derive the relations found in Ref.~\cite{Hansen:2015zga}, we will not replace $\Gc$ by $\Gc\,'$ everywhere, and instead keep the original $\Gc$ in the definition of $\Dc^{(u,u)}$ as given in Eq.~\eqref{eq:ladder}. While this may seem awkward at first, we can rationalize it by following the same steps in Sec.~\ref{sec:alternative_derivation} until we reach the unitarity equation~\eqref{eq:M3df_unitarity} for $\Mc_{3,\df}^{(u,u)}$, and as this point choose the new $\Gc\,'$ to arrive at the relations for $\Tc_{\df}^{(u,u)\,\prime}$, 
\begin{subequations}
	\begin{align}
	\Mc_{3,\df}^{(u,u)}(\p,\k) & = \int_{p'}\int_{k'} \, \Lc^{(u,u)\,\prime}(\p,\p')\, \Tc_{\df}^{(u,u)\, \prime}(\p',\k')\,\Rc^{(u,u)\,\prime}(\k',\k)  \, , \\[5pt]
	\label{eq:Lfcn_prime}
	\Lc^{(u,u)\,\prime}(\p,\k) & =  \delta(\p,\k) - \int_{k'}  \Big[ \, \delta(\p,\k') \, \Mc_{2}(\sigma_{k'}) + \Dc^{(u,u)}(\p,\k') \, \Big] \, \Gamma\,'(\k',\k) \, ,\\[5pt]
	\label{eq:Rfcn_prime}
	\Rc^{(u,u)\,\prime}(\p,\k) & = \delta(\p,\k) - \int_{k'} \Gamma\,'(\p,\k') \, \Big[ \, \delta(\k',\k) \, \Mc_{2}(\sigma_k) + \Dc^{(u,u)}(\k',\k) \, \Big] \, , \\[5pt]
	\label{eq:Tdf_onshell_prime}
	\Tc_{\df}^{(u,u)\,\prime}(\p,\k) & = \Kc_{3,\df}^{(u,u)\,\prime}(\p,\k) \nn \\[5pt]
	& - \int_{p'}\int_{q'}\int_{k'} \Kc_{3,\df}^{(u,u)\,\prime}(\p,\p') \Gamma\,'(\p',\q') \Lc^{(u,u)\,\prime}(\q',\k') \Tc_{\df}^{(u,u)\,\prime}(\k',\k) \, ,
	\end{align}
\end{subequations}
where again $\Dc^{(u,u,)}$ is given by Eq.~\eqref{eq:ladder}. This procedure still leads to a consistent on-shell representation with respect to unitarity, and will be the basis for the following manipulations.

It is convenient to express $\Tc_{\df}^{(u,u)\,\prime}$ as its series in the number of $\Kc_{3,\df}^{(u,u)\,\prime}$ insertions,
\begin{align}
	\Tc_{\df}^{(u,u)\,\prime}(\p,\k) = \sum_{n = 1}^{\infty} \Tc_{\df, n}^{(u,u)\,\prime}(\p,\k) \, , \nn 
\end{align}
where the leading and next-to-leading order terms are
\begin{align}
	\Tc_{\df, 0}^{(u,u)\,\prime}(\p,\k) & = \Kc_{3,\df}^{(u,u)\,\prime}(\p,\k) \, , \nn \\[5pt]
	\Tc_{\df, 1}^{(u,u)\,\prime}(\p,\k) & = - \int_{p'}\int_{q'}\int_{k'} \Kc_{3,\df}^{(u,u)\,\prime}(\p,\p') \Gamma\,'(\p',\q') \Lc^{(u,u)\,\prime}(\q',\k') \Kc_{3,\df}^{(u,u)\,\prime}(\k',\k) \, , \nn 
\end{align}
where all other orders follow from successive substitutions of Eq.~\eqref{eq:Tdf_onshell_prime} into itself. For our purposes, the lowest two orders provide the structure needed to generalize to the infinite series generated by Eq.~\eqref{eq:Tdf_onshell_prime}.

At leading order in $\Kc_{3,\df}^{(u,u)\,\prime}$, $\Mc_{3,\df}^{(u,u)}$ has the form
\begin{align}
	\label{eq:M3df_LO}
	\Mc_{3,\df}^{(u,u)}(\p,\k) \Big\rvert_{\mathrm{LO}} =  \int_{p'}\int_{k'} \, \Lc^{(u,u)\,\prime}(\p,\p')\, \Kc_{3,\df}^{(u,u)\,\prime}(\p',\k')\,\Rc^{(u,u)\,\prime}(\k',\k) \, .
\end{align}
Focusing on the operation of $\Rc^{(u,u)\,\prime}$ on the initial state, \ie the expression
\begin{align}
	\label{eq:symm_initial}
	\int_{p'}\Kc_{3,\df}^{(u,u)\,\prime}(\p,\p')\, & \Rc^{(u,u)\,\prime}(\p',\k) = \Kc_{3,\df}^{(u,u)\,\prime}(\p,\k) \nn \\
	& - \int_{p'}\int_{k'} \Kc_{3,\df}^{(u,u)\,\prime}(\p,\p')\, \Gamma\,'(\p',\k') \, \left[ \, \delta(\k',\k) \,\Mc_{2}(\sigma_k) + \Dc^{(u,u)}(\k',\k) \, \right] \, ,
\end{align}
the goal is to use properties of the symmetrization and exchange cut to simplify this expression and cast it into a form where the $K$ matrix is symmetric under interchange of particles. Two identities are used, which are derived in Refs.~\cite{Hansen:2014eka,Hansen:2015zga,Blanton:2020gha} and summarized in Appendix \ref{sec:app_sym}, to convert the $K$ matrix from the unsymmetrized to a symmetrized form. For the first term of Eq.~\eqref{eq:symm_initial}, we note that the initial state ultimately would be summed over all partial waves and symmetrized over all spectators, 
\begin{align}
	\Kc_{3,\df}^{(u)}(\p,\k,\a_k^{\star}) = \sqrt{4\pi}\sum_{\k\in\Pc_k} \sum_{\ell,m_{\ell}} Y_{\ell m_{\ell}}^{*}(\bh{\a}_k^{\star}) \, \Kc_{3,\df}^{(u,u)}(\p,\k) \, ,
\end{align}
where the left-hand side is only asymmetric in the final state ($\p$). Since all the spectators are involved on the right-hand side, we can use the identity Eq.~\eqref{eq:app_sym.Xsymm_final} to replace the $K$ matrix as
\begin{align}
	\Kc_{3,\df}^{(u,u)\,\prime}(\p,\k) \longrightarrow \frac{1}{3} \, \left[ \,  \Kc_{3,\df}^{(u,u)\,\prime}(\p,\k)  + 2\Kc_{3,\df}^{(u,s)\,\prime}(\p,\k) \,\right] \, , \nn
\end{align}
where $\Kc_{3,\df}^{(u,s)\,\prime}$ are the projected $K$ matrices in the non-$\k$ spectators, and the factor of 2 is because both spectators contribute identically since identical scalar particles only have even partial waves contributions, \cf Eqs.~\eqref{eq:app_sym.Xsymm} and \eqref{eq:app_sym.Xcomb}. The $1/3$ factor arises from overcounting the initial state as we sum over all spectators, see Eq.~\eqref{eq:app_symm.Xo3}. 

The second term of Eq.~\eqref{eq:symm_initial} can be symmetrized as well. We use the identity~\eqref{eq:app_sym.identity2} which relates the exchange $\Gc'$ to $\Ic$ when summing over all angular momenta and integrating over the entire spectator momentum space against a short-distance function. We find that for the initial state
\begin{align}
	\int_{k'} \Kc_{3,\df}^{(u,u)\,\prime}(\p,\k') \Gamma\,'(\k',\k) & = \int_{k'} \Kc_{3,\df}^{(u,u)\,\prime}(\p,\k') \left[ \, \delta(\k',\k)\,\wt{\rho}(\sigma_k) + \Gc\,'(\k',\k)  \, \right] \, , \nn \\[5pt]
	& = \left[ \, \Kc_{3,\df}^{(u,u)\,\prime} (\p,\k) + 2\Kc_{3,\df}^{\prime\,(u,s)} (\p,\k)\, \right] \, \wt{\rho}(\sigma_k) \, , \nn
\end{align}
where we have defined $\wt{\rho}(\sigma_k) \equiv H(\sigma_k) \, \Ic(\sigma_k)$. 

Note that so far we have only symmetrized the initial state, we can apply the same identities for the final state of Eq.~\eqref{eq:M3df_LO}. We then define a \emph{symmetric} $\Kc_{3,\df}$ as
\begin{align}
	\Kc_{3,\df}(\p,\k) & \equiv \Kc_{3,\df}^{(u,u)\,\prime}(\p,\k) +2 \Kc_{3,\df}^{(u,s)\,\prime}(\p,\k)  + 2  \Kc_{3,\df}^{(s,u)\prime}(\p,\k)  + 4 \Kc_{3,\df}^{(s,s)\prime}(\p,\k) \, .
\end{align}
Then the leading order $\Mc_{3,\df}$ is can equivalently be written as
\begin{align}
	\Mc_{3,\df}^{(u,u)}(\p,\k) \Big\rvert_{\mathrm{LO}} = \int_{p'}\int_{k'} \Lc(\p,\p') \, \Kc_{3,\df}(\p',\k') \, \Rc(\k',\k)  \, , \nn 
\end{align}
where we have defined new symmetric initial and final state rescattering functions
\begin{align}
	\Lc(\p,\k) & = \left[ \, \frac{1}{3}  -  \Mc_{2}(\sigma_k) \, \wt{\rho}(\sigma_k) \, \right]  \delta(\p,\k) - \Dc^{(u,u)}(\p,\k) \wt{\rho}(\sigma_k) \, , \nn \\
	\Rc(\p,\k) & = \left[ \, \frac{1}{3}  - \wt{\rho}(\sigma_p) \, \Mc_{2}(\sigma_p) \, \right]  \delta(\p,\k) - \wt{\rho}(\sigma_p) \, \Dc^{(u,u)}(\p,\k) \, . \nn
\end{align}
The next-to-leading order term of $\Mc_{3,\df}^{(u,u)}$ contains two occurrences of $\Kc_{3,\df}^{(u,u)}$, both of which are dressed on the external legs by $\Lc$ and $\Rc$ in the same way as the leading order. We have at this order
\begin{align}
	\Mc_{3,\df}^{(u,u)}(\p,\k) \Big\rvert_{\mathrm{NLO}} & = \int_{p'}\int_{k'} \, \Lc^{(u,u)\,\prime}(\p,\p')\, \Tc_{\df,1}^{(u,u)\,\prime}(\p',\k')\,\Rc^{(u,u)\,\prime}(\k',\k)  \, ,\nn \\[5pt]
	& = -\int_{p'}\int_{q''} \int_{q'}\int_{q} \int_{k'} \, \Lc^{(u,u)\,\prime}(\p,\p')\, \Kc_{3,\df}^{(u,u)\,\prime}(\p',\q'') \, \Gamma \,'(\q'',\q') \nn \\[5pt]
	& \qquad \times \, \Lc^{(u,u)\,\prime}(\q',\q) \, \Kc_{3,\df}^{(u,u)\,\prime}(\q,\k') \,\Rc^{(u,u)\,\prime}(\k',\k) \, ,
\end{align}
which we can apply the same identities to both $K$ matrices. We find that we can rewrite the next-to-leading order term in the symmetric form as
\begin{align}
	\Mc_{3,\df}^{(u,u)}(\p,\k) \Big\rvert_{\mathrm{NLO}} & = -\int_{p'}\int_{q'} \int_{q} \int_{k} \, \Lc(\p,\p')\, \Kc_{3,\df}(\p',\q') \, \wt{\rho}(\sigma_{q'}) \nn \\[5pt]
	& \qquad \times \, \Lc(\q',\q) \Kc_{3,\df}(\q,\k') \,\Rc(\k',\k) \, , \nn \\[5pt]
	& \equiv \int_{p'}\int_{k'} \Lc(\p,\p') \, \Tc_{\df,1}(\p',\k') \, \Rc(\k',\k) \, ,\nn
\end{align}
where $\Lc$ and $\Rc$ are as above and we have defined $\Tc_{\df}$ as
\begin{align}
	\Tc_{\df}(\p,\k) = - \int_{p'}\int_{k'}  \Kc_{3,\df}(\p,\p') \, \wt{\rho}(\sigma_{p'}) \, \Lc(\p',\k') \Kc_{3,\df}(\k',\k') \, . \nn
\end{align}

After applying these operations on all occurrences of $\Kc_{3,\df}$ in $\Mc_{3,\df}^{(u,u)}$, the symmetric $K$ matrix form of the $\Mc_{3,\df}^{(u,u)}$ equation reads
\begin{align}
\label{eq:amputate_symm}
\Mc_{3,\df}^{(u,u)}(\p,\k) \equiv \int_{p'}\int_{k'} \, \Lc(\p,\p')\, \Tc_{\df}(\p',\k')\,\Rc(\k',\k) \, ,
\end{align}
where $\Tc_{\df}$ is a symmetric amputated amplitude satisfying the linear integral equation driven by the symmetric $K$ matrix
\begin{align}
	\label{eq:Tdf_symm_onshell}
	\Tc_{\df}(\p,\k) = \Kc_{3,\df}(\p,\k) - \int_{p'}  \int_{k'} \, \Kc_{3,\df}(\p,\p') \, \wt{\rho}(\sigma_{p'}) \, \Lc(\p',\k') \, \Tc_{\df}(\k',\k) \, .
\end{align}
Equation~\eqref{eq:Tdf_symm_onshell} is identical to the one derived in Ref.~\cite{Hansen:2015zga} provided the cutoff functions are chosen to be the same. We can combine Eq.~\eqref{eq:Tdf_symm_onshell}, \eqref{eq:amputate_symm}, and \eqref{eq:ladder} into a single integral equation similar to its asymmetric form~\eqref{eq:M3_onshell}. Alternatively, we can arrive at the same relation by applying the symmetry operations to Eq.~\eqref{eq:M3_onshell} directly. We find that the connected $\3\to\3$ amplitude satisfies the integral equation
\begin{align}
	\Mc_3^{(u,u)}(\p,\k) & = \frac{1}{9} \, \Kc_{3,\df}(\p,\k) - \Kc_2(\sigma_{p}) \, \Gc(\p,\k) \, \Mc_2(\sigma_k) - \frac{1}{3} \Kc_{3,\df}(\p,\k) \, \wt{\rho}(\sigma_k) \, \Mc_{2}(\sigma_k) \, \nn \\[5pt]
	& \qquad - \Kc_2(\sigma_p) \, \int_{k'}\,\Gamma(\p,\k') \, \Mc_{3}^{(u,u)}(\k',\k) - \frac{1}{3}\int_{k'} \Kc_{\df,3}(\p,\k') \, \wt{\rho}(\sigma_{k'}) \, \Mc_{3}^{(u,u)}(\k',\k) \, ,
\end{align}
where $\Gamma(\p,\k) = \delta(\p,\k) \, H(\sigma_k) \, \Ic(\sigma_k) + \Gc(\p,\k) \equiv \delta(\p,\k) \,\wt{\rho}(\sigma_k) + \Gc(\p,\k)$ as before. We have reduced the original set of integral equations from Ref.~\cite{Hansen:2015zga} to a simple single integral equation, where we calculated $\Mc_{3}^{(u,u)}$ directly given $\Kc_2$ and $\Kc_{3,\df}$. The numerical factors arise from over-counting the states as we write the relation in terms of the unsymmetrized $\Mc_{3}^{(u,u)}$.

As a final remark, the finite-volume equivalents can be found by following the same arguments in the beginning of Sec.~\ref{sec:alternative}, which we simply state the result
\begin{align}
	\det \Big[ \, 1 + \Kc_{3,\df}(P) \cdot F_{3,L}(P) \,\Big]_{E = E_{\mathfrak{n}} }= 0 \, ,
\end{align}
where $F_{3,L}$ is a symmetrized version of $F_{3,L}^{(u,u)}$ introduced earlier,
\begin{align}
	F_{3,L}(P) \equiv \wt{F}_{2,L} \cdot \left( \, \frac{1}{3} - \frac{1}{{\wt{\Kc}_{2,L}}^{\, -1} + \wt{F}_{2,L} + \wt{G}_{L}(P)} \cdot \wt{F}_{2,L} \right) \, , \nn 
\end{align}
which agrees with the original quantization condition presented in Ref.~\cite{Hansen:2014eka}.

\section{Conclusions}
\label{sec:conclusions}

We have presented a general way to derive relativistic $\3\to\3$ scattering integral equations and finite volume quantization conditions using $S$ matrix unitarity. The method uses the fact that singularities from on-shell physics is responsible for the dominant finite-volume effects and $S$ matrix unitarity constrains all physical singularities in the kinematic region of interest. The method is non-perturbative, and considers only on-shell amplitudes, avoiding the need to connect these quantities to quantum fields. This approach differs from previous FVU studies~\cite{Mai:2017vot,Mai:2017bge} by considering the $T$ matrix elements directly, which give a simpler path toward both the scattering equations and the finite-volume quantization conditions.

While the methods were shown to be equivalent~\cite{Jackura:2019bmu,Blanton:2020jnm}, this work provides a direct connection between the RFT and FVU approaches. We found that the TOPT version of the RFT approach is most directly related to the FVU method, which was noted in Ref.~\cite{Blanton:2020gha} in their derivation of the quantization conditions. The connection manifest itself by examining the on-shell representation of the $T$ matrix elements, not the fully connected amplitude, and embedding the system in a confined volume. Making this connection more explicit allowed us to find a new expression for the $\3\to\3$ $T$ matrix integral equation shown in Eq.~\eqref{eq:Tuu_onshell}, as well as gave new physical insight the $\Tc_{\df}$ object which can be interpreted as the kinematically singular free $T$ matrix element which was shown in Sec.~\ref{sec:alternative}. We were able to recover the quantization conditions of Ref.~\cite{Blanton:2020gha} directly, shown in Eq.~\eqref{eq:QC3_v1}. By using various identities and manipulations, we were able to transform the simple on-shell form for the $T$ matrix element and the resulting quantization conditions to all previously derived results for three identical particles, including new versions for the asymmetric three-body $K$ matrix given in Eq.~\eqref{eq:M3_onshell} for the amplitude and Eq.~\eqref{eq:QC3_new} for the associated quantization condition.

Our approach here is rather efficient in generating the scattering equations and quantization conditions, requiring only that unitarity condition for the $T$ matrix element and the kinematic functions which characterize the on-shell propagation of intermediate state particles. The results can easily be generated to particles of different species, coupled two and three channel systems, and particles with arbitrary spin once the state space is fully defined. Moreover, we find this direction promising for extending the framework to higher number of particles, \eg four particle systems. Combined with the recent efficient version of the all-orders approach using TOPT~\cite{Blanton:2020gha}, this framework provides a unified approach to linking lattice QCD observables to few-body phenomena.

\section*{Acknowledgements}
The author would like to thank R.~Brice\~no, S.~Dawid, and F.~Romero-L\'opez for helpful discussions and comments on the manuscript, and would also like to thank M.~Hansen, S.~Sharpe, and A.~Szczepaniak for useful discussions regarding this work. A.W.J. acknowledges supported from U.S. Department of Energy contract DE-AC05-06OR23177, under which Jefferson Science Associates, LLC, manages and operates Jefferson Lab. A.W.J. also acknowledges support of the USDOE Early Career award, contract DE-SC0019229.

\appendix

\section{Three pariticle unitarity relation}
\label{sec:app_unitarity}

In this appendix, we provide some details regarding the derivation of Eq.~\eqref{eq:Tuu_unitarity}, which has been presented in Ref.~\cite{Jackura:2018xnx}, with some modifications to agree with the conventions set in this manuscript. Since we are interested in amplitudes within the pair-spectator basis, we find immediately for these states that the unitarity relation has the form
\begin{align}
	\label{eq:app.unitarity.unitarity1}
	2\im \bra{P,\p,\ell' m_{\ell'}} T \ket{P,\k,\ell m_{\ell}} = \bra{P,\p,\ell' m_{\ell'}} T^{\dag} T\ket{P,\k,\ell m_{\ell}} \, ,
\end{align}
where Hermitian analyticity is assumed to write the left hand side as the imaginary part of the amplitude~\cite{Olive:1962,Eden:1966dnq}. We insert the identity between $T^{\dag} T$ in the form of the completeness relation for identical three particle states,
\begin{align}
	\label{eq:app.unitarity.completeness}
	\mathbbm{1} = \frac{1}{3!} \int_{k_1} \int_{k_2} \int_{k_3} \, \ket{\k_1,\k_2,\k_3} \! \bra{\k_1,\k_2,\k_3} \, ,
\end{align}
where the division by $3!$ is the symmetry factor since the particles are identical. No other multiparticle states contribute in the completeness relation since energies in the elastic three particle scattering region are considered only.

In terms of the pair-spectator basis~\eqref{eq:state}, the identity splits into two distinct topologies: one where the spectator is remains the same in the intermediate state, and one where it exchanges
\begin{align}
	\label{eq:app.unitarity.completeness_pair_spectator}
	\mathbbm{1} & = \frac{1}{2!}\sum_{\ell,m_{\ell}} \int_{k} \int_{0}^{\infty} \! \diff p_k^{\star} \, \frac{p_k^{\star\,2}}{(2\pi)^2 \, \omega_{p_k}^{\star}} \int_{P}  \, \ket{P,\k,\ell m_{\ell}} \! \bra{P,\k,\ell m_{\ell}} \nn \\[5pt]
	& \qquad + \sum_{\ell',m_{\ell'}} \sum_{\ell,m_{\ell}} \int_{k}\int_{p} \int_{P} \, 4\pi Y_{\ell' m_{\ell'}}(\bh{\k}_p^{\star}) Y_{\ell m_{\ell}}^{*}(\bh{\p}_k^{\star}) \, \ket{P,\p,\ell' m_{\ell'}} \! \bra{P,\k,\ell m_{\ell}} \, .
\end{align}
The first term arises when the spectators in both states are equal, giving an overall factor of 3 since the particles are identical and leaving a factor $1/2!$ which is the symmetry factor for the two particles composing the pair. Additionally, the measure is transformed into the pair CM frame, \ie the frame where $\P^{\star} - \k^{\star} = \0$, which integrates away the angular contribution via the spherical harmonic orthonormality relation leaving the magnitude of the relative momentum of the pair $p^{\star}$. The shorthand for integral labeled by $P$ follows the definition
\begin{align}
	\label{eq:app.unitarity.pmom}
	\int_P \equiv \int \! \frac{\diff^3 \P}{(2\pi)^3 \, 2\omega_{Pkp}} \, ,
\end{align}
with $\omega_{Pkp} = \sqrt{m^2 + \lvert \P - \k-\p \rvert^2}$. Note for the first term of Eq.~\eqref{eq:app.unitarity.completeness_pair_spectator}, we have boosted to the $\P^{\star} - \k^{\star} = \0$ frame, so the energy factor in Eq.~\eqref{eq:app.unitarity.pmom} is $\omega_{Pkp} = \sqrt{m^2 + \lvert \p_k^{\star} \rvert^2}$. In the second term, the spectators $\p$ and $\k$ are not the same, which leaves 6 identical terms that cancels the symmetry factor.

Inserting Eq.~\eqref{eq:app.unitarity.completeness_pair_spectator} into \eqref{eq:app.unitarity.unitarity1}, along with the definition of the asymmetric amplitude Eq.~\eqref{eq:Tmatrix_symm}, we find the unitarity relation
\begin{align}
	\label{eq:app.unitarity.unitarity2}
	2\im \Tc_{\ell' m_{\ell'} , \ell m_{\ell}}^{(u,u)}(\p,\k) & =  \frac{1}{2!}\sum_{\ell'',m_{\ell''}}\int_{k'}  \int_{0}^{\infty} \! \diff p_{k'}'^{\star} \, \frac{p_{k'}'^{\star\,2}}{(2\pi)^2 \, \omega_{p_{k'}'}^{\star}}  \int_{P'} \,(2\pi)^{4} \delta^{(4)}(P'-P) \nn \\
	& \qquad \times \Tc_{\ell'' m_{\ell''} , \ell' m_{\ell'}}^{(u,u)\,*}(\p,\k') \Tc_{\ell'' m_{\ell''} , \ell m_{\ell}}^{(u,u)}(\k',\k) \nn \\
	& + \sum_{\ell''',m_{\ell'''}} \sum_{\ell'',m_{\ell''}} \int_{k'}\int_{p'}  \int_{P'}  \, 4\pi Y_{\ell''' m_{\ell'''}}(\bh{\k}_{p'}'^{\star}) Y_{\ell'' m_{\ell''}}^{*}(\bh{\p}_{k'}'^{\star}) \, (2\pi)^4\delta^{(4)} (P'-P) \nn \\
	& \qquad \times \Tc_{\ell''' m_{\ell'''} , \ell' m_{\ell'}}^{(u,u)\,*}(\p,\p') \Tc_{\ell'' m_{\ell''} , \ell m_{\ell}}^{(u,u)}(\k',\k)  \, .
\end{align}
The integration over $\P'$ in both terms is trivially evaluated using $(2\pi)^3\,\delta^{(3)}(\P'-\P)$, leaving a single delta function over the energy. Since the spectator energy is conserved in the first term, the remaining delta function has an argument $E'-E = (E'-\omega_{k'}) - (E - \omega_{k'})$, and therefore the integral over $p_{k'}'^{\star}$ evaluates to $2\rho(\sigma_{k'})$ given in Eq.~\eqref{eq:direct_cut} since the amplitudes are independent of $p_{k'}'^{\star}$. The energy delta function in the second term of Eq.~\eqref{eq:app.unitarity.unitarity2} is most readily evaluated by using the identity
\begin{align}
	\frac{1}{2\omega_{Pk'p'}} \delta(E - \omega_{p'} - \omega_{k'} - \omega_{Pk'p'}) = \delta((P-k'-p')^2 - m^2) \Theta(E-\omega_{k'}-\omega_{p'}) \, , \nn
\end{align}
which reduces to $2 \Delta(\p,\k)$ since the Heaviside function enforces that the intermediate state is a physical process, which is equivalent to the definition given in Eq.~\eqref{eq:exchange_cut}. Given these evaluations, Eq.~\eqref{eq:app.unitarity.unitarity2} reduces to \eqref{eq:Tuu_unitarity}, as claimed in Sec.~\ref{sec:scattering}.

\section{Proof of unitarity for scattering equations}
\label{sec:app_proof}

We show how the various \ansatz for the $\3\to\3$ scattering equations satisfy the unitarity relations. The approach presented here is more efficient than those presented in Refs.~\cite{Jackura:2018xnx,Briceno:2019muc}. It is instructive to illustrate the method by demonstrating that the simpler $\2\to\2$ on-shell representation satisfies Eq.~\eqref{eq:M2_unitarity}. First, we take the Hermitian conjugate of Eq.~\eqref{eq:M2_onshell}, and use $\Kc_2^{\dag} = \Kc_{2}$ in the physical region to find
\begin{align}
	\label{eq:app_K2}
	\Kc_{2} = \Mc_{2}^{\dag} + \Mc_{2}^{\dag} \, \Ic^{\dag} \, \Kc_{2} \, .
\end{align}
The, we substitute Eq.~\eqref{eq:app_K2} into the second term of Eq.~\eqref{eq:M2_onshell} to find
\begin{align}
	\label{eq:app_M2}
	\Mc_{2} = \Kc_{2} - \Mc_{2}^{\dag} \, \Ic \, \Mc_{2} - \Mc^{\dag} \, \Ic^{\dag} \, \Kc_{2} \, \Ic \, \Mc_{2} \, .
\end{align}
Finally, we take the Hermitian conjugate of Eq.~\eqref{eq:app_M2}, and subtract this from~\eqref{eq:app_M2} to find
\begin{align}
	\Mc_{2} - \Mc_{2}^{\dag} = - \Mc_{2}^{\dag} \, \left[ \, \Ic - \Ic^{\dag} \,\right] \, \Mc_{2} \, , 
\end{align}
which with $\Ic - \Ic^{\dag} = 2i \, \im\Ic = -2i\, \rho$ and $\Mc_{2} - \Mc_{2}^{\dag} = 2i \, \im \Mc_{2}$ since $\Mc_{2}^{\top} = \Mc_{2}$ by time-reversal invariance, recovers Eq.~\eqref{eq:M2_unitarity} and completes the demonstration.

We follow the same strategy for the $\3\to\3$ amplitude, applying it to the integral equation~\eqref{eq:Tuu_onshell} for the full $T$ matrix element, given here for convenience
\begin{align}
	\Tc^{(u,u)}(\p,\k) = \Kc_{3}^{(u,u)}(\p,\k) - \int_{p'} \int_{k'} \Kc_{3}^{(u,u)}(\p,\p') \,  \Gamma(\p',\k')  \, \Tc^{(u,u)}(\k',\k) \, . \nn 
\end{align}
Note that while there is a cluster decomposition in $\Tc^{(u,u)}$ and $\Kc_3^{(u,u)}$, Eqs.~\eqref{eq:Tuu_connected} and \eqref{eq:Kmat_connected} respectively, it does not spoil the proof. Taking now the Hermitian conjugate and using the fact that the $K$ matrices are real symmetric matrices, we find
\begin{align}
	\Kc_{3}^{(u,u)}(\p,\k) = \Tc^{(u,u)\,\dag}(\p,\k) + \int_{p'} \int_{k'} \Tc^{(u,u)\,\dag}(\p,\p') \,  \Gamma^{\dag}(\p',\k')  \, \Kc_{3}^{(u,u)}(\k',\k) \, .
\end{align}
This equation is substituted into the second term of Eq.~\eqref{eq:Tuu_onshell}, giving
\begin{align}
	\Tc^{(u,u)}(\p,\k) & = \Kc_{3}^{(u,u)}(\p,\k) - \int_{p'} \int_{k'} \Tc^{(u,u)\,\dag}(\p,\p') \,  \Gamma(\p',\k')  \, \Tc^{(u,u)}(\k',\k) \, , \nn \\[5pt]
	& - \int_{p'} \int_{q'} \int_{q} \int_{k'} \Tc^{(u,u)\,\dag}(\p,\p') \,  \Gamma^{\dag}(\p',\q')  \, \Kc_{3}^{(u,u)}(\q',\q) \,  \Gamma(\q,\k')  \, \Tc^{(u,u)}(\k',\k) \, .
\end{align}
We can now take the difference between $\Tc^{(u,u)}$ and its Hermitian conjugate to arrive at
\begin{align}
	\Tc^{(u,u)}(\p,\k) - \Tc^{(u,u)\,\dag}(\p,\k) = - \int_{p'} \int_{k'} \Tc^{(u,u)\,\dag}(\p,\p') \,  \left[ \, \Gamma(\p',\k') - \Gamma^{\dag}(\p',\k') \, \right] \, \Tc^{(u,u)}(\k',\k) \, , \nn
\end{align}
which is the unitarity relation~\eqref{eq:Tuu_unitarity} since $\Tc^{(u,u)} - \Tc^{(u,u)\,\dag} = 2i \, \im \Tc$ and $\Gamma - \Gamma^{\dag} = 2i \, \im \Gamma = - 2i \Phi$. We conclude that the linear integral equation of the form~\eqref{eq:Tuu_onshell} is a valid on-shel representation. This method works similarly to the other forms of the integral equations, such as for $\Tc_{\df}^{(u,u)}$ in Eq.~\eqref{eq:Tdf_onshell}.

\section{Relating $K$ matrices}
\label{sec:app_K_matrices}

In this appendix we outline the procedure for relating two different $K$ matrices defined by their respective kinematic functions. We work with the integral equation~\eqref{eq:Tuu_onshell} for $\Tc^{(u,u)}$, with one scheme defined by
\begin{align}
	\label{eq:app_relate.K}
	\Tc^{(u,u)}(\p,\k) = \Kc_{3}^{(u,u)}(\p,\k) - \int_{p'} \int_{k'} \Kc_{3}^{(u,u)}(\p,\p') \,  \Gamma(\p',\k')  \, \Tc^{(u,u)}(\k',\k) \, , 
\end{align}
and the other defined with
\begin{align}
	\label{eq:app_relate.Kprime}
	\Tc^{(u,u)}(\p,\k) = \Kc_{3}^{(u,u)\,\prime}(\p,\k) - \int_{p'} \int_{k'} \Tc^{(u,u)}(\p,\p') \,  \Gamma\,'(\p',\k')  \, \Kc_{3}^{(u,u)\,\prime}(\k',\k) \, , 
\end{align}
where $\Gamma' = \Gamma + \delta \Gamma'$ with $\delta\Gamma\,'$ being the difference in the schemes, and we have reordered the second equation which can be seen by expanding the Fredholm series and resumming on the final state. Note that both the disconnected and connected $K$ matrices may be affected by the scheme, \ie $\Kc_3^{(u,u)\,\prime} = \delta(\p,\k)\,\Kc_2\,' + \Kc_{3,\df}^{(u,u)\,\prime}$. We now form two new equations, one by substituting Eq.~\eqref{eq:app_relate.Kprime} into the second term of \eqref{eq:app_relate.K},
\begin{align}
	\label{eq:app_relate.K2}
	\Tc^{(u,u)}(\p,\k) & = \Kc_{3}^{(u,u)}(\p,\k) - \int_{p'} \int_{k'} \Kc_{3}^{(u,u)}(\p,\p') \,  \Gamma(\p',\k')  \, \Kc_{3}^{(u,u)\,\prime}(\k',\k) \, \nn \\[5pt]
	& + \int_{p'} \int_{q'} \int_{q}  \int_{k'} \Kc_{3}^{(u,u)}(\p,\p') \,  \Gamma(\p',\q') \Tc^{(u,u)}(\q',\q) \,  \Gamma\,'(\q,\k')  \, \Kc_{3}^{(u,u)\,\prime}(\k',\k)
\end{align}
and the second by substituting~\eqref{eq:app_relate.K} into the second term of \eqref{eq:app_relate.Kprime},
\begin{align}
	\label{eq:app_relate.Kprime2}
	\Tc^{(u,u)}(\p,\k) & = \Kc_{3}^{(u,u)\,\prime}(\p,\k) - \int_{p'} \int_{k'} \Kc_{3}^{(u,u)}(\p,\p') \,  \Gamma\,'(\p',\k')  \, \Kc_{3}^{(u,u)\,\prime}(\k',\k) \, \nn \\[5pt]
	& + \int_{p'} \int_{q'} \int_{q}  \int_{k'} \Kc_{3}^{(u,u)}(\p,\p') \,  \Gamma(\p',\q') \Tc^{(u,u)}(\q',\q) \,  \Gamma\,'(\q,\k')  \, \Kc_{3}^{(u,u)\,\prime}(\k',\k) \, .
\end{align}
Notice that the last term in Eqs.~\eqref{eq:app_relate.Kprime2} and \eqref{eq:app_relate.K2} are identical. We can now subtract Eq.~\eqref{eq:app_relate.K2} from \eqref{eq:app_relate.Kprime2}, giving the relation
\begin{align}
	\label{eq:app_relate.Krelate}
	\Kc_{3}^{(u,u)}(\p,\k) = \Kc_{3}^{(u,u)\,\prime}(\p,\k) - \int_{p'} \int_{k'} \Kc_{3}^{(u,u)}(\p,\p') \,  \delta\Gamma\,'(\p',\k')  \, \Kc_{3}^{(u,u)\,\prime}(\k',\k) \, ,
\end{align}
where we have used $\Gamma\,' = \Gamma + \delta \Gamma\,'$. Given two different schemes, the master equation~\eqref{eq:app_relate.Krelate} relates the resulting $K$ matrices.

As an example, let's consider $\delta\Gamma\,' = - \Gc_{\pv}$ as in Sec.~\ref{sec:symmetric_K}. In this case, $\Kc_2$ will be unchanged, and the only difference is $\Kc_{3,\df}^{(u,u)}$. We can manipulate the master equation~\eqref{eq:app_relate.Krelate} into a form for just $\Kc_{3,\df}^{(u,u)}$. Using  $\Kc_3^{(u,u)\,\prime} = \delta(\p,\k)\,\Kc_2 + \Kc_{3,\df}^{(u,u)\,\prime}$, we find
\begin{align}
	\Kc_{3,\df}^{(u,u)}(\p,\k) & = \Kc_{3,\df}^{(u,u)\,\prime}(\p,\k)  + \Kc_2(\sigma_p) \,  \Gc_{\pv}(\p,\k)  \, \Kc_2(\sigma_k)   \nn \\[5pt]
	& + \int_{k'}  \Kc_2(\sigma_p)  \,  \Gc_{\pv}(\p,\k')  \,  \Kc_{3,\df}^{(u,u)\,\prime}(\k',\k)  + \int_{p'}  \Kc_{3,\df}^{(u,u)}(\p,\p')  \,  \Gc_{\pv}(\p',\k)  \, \Kc_2(\sigma_k)   \nn \\[5pt]
	& + \int_{p'} \int_{k'} \Kc_{3,\df}^{(u,u)}(\p,\p') \,  \Gc_{\pv}(\p',\k')  \, \Kc_{3,\df}^{(u,u)\,\prime}(\k',\k) \, . \nn 
\end{align}
This relation can be reorganized into a more compact form as
\begin{align}
	\Kc_{3,\df}^{(u,u)}(\p,\k) & = \int_{p'} \int_{k'} \Cc_L(\p,\p') \, \Kc_{3,\df}^{(u,u) \, \prime}(\p',\k') \, \Cc_R(\k',\k) \nn \\
	& + \int_{p'}\int_{q'}\int_{k'} \Cc_L(\p,\p') \, \Kc_{3,\df}^{(u,u)\, \prime}(\p',\q')  \Gc_{\pv}(\q',\k') \, \Kc_{3,\df}^{(u,u)}(\k',\k) \, , \nn
\end{align}
where
\begin{align}
	\Cc_L(\p,\k) & = \delta(\p,\k) + \Kc_2(\sigma_p) \, \Gc_{\pv}(\p,\k) \, , \nn \\[5pt]
	\Cc_R(\p,\k) & = \delta(\p,\k) +  \Gc_{\pv}(\p,\k) \, \Kc_2(\sigma_k) \, . \nn
\end{align}
which is the result presented in Eq.~\eqref{eq:Kmat_relation} in Sec.~\ref{sec:symmetric_K}.

\section{Symmetrization identities}
\label{sec:app_sym}

This appendix serves to summarize important properties of the asymmetric amplitudes, which have been detailed in Refs.~\cite{Hansen:2014eka,Hansen:2015zga,Blanton:2020gha}. For ease of demonstration, let us consider a generic function $X(\k,\a)$ which suppresses the momenta of the final state, and $\k$, $\a$, and $\b = \P-\k-\a$ are the momenta of the initial three particle state. From the symmetrization definition Eq.~\eqref{eq:Tmatrix_symm}, the asymmetric amplitude is defined as
\begin{align}
	\label{eq:app_sym.def}
	X(\k,\a) & \equiv \sum_{\k \in \{\k,\a,\b\}} X^{(u)}(\k,\bh{\a}_k^{\star}) \, , \nn \\
	& = X^{(u)}(\k,\bh{\a}_k^{\star}) + X^{(u)}(\a,\bh{\k}_a^{\star}) + X^{(u)}(\b,\bh{\k}_b^{\star}) \, ,
\end{align}
The $(u)$ superscript indicates that the amplitude is asymmetric since the spectator momentum is distinguished. In addition to this definition, Ref.~\cite{Hansen:2014eka} introduced an additional notion where the spectator is labeled with alternative indices $s$ and $\wt{s}$, \ie
\begin{subequations}
	\begin{align}
	\label{eq:app_sym.s}
	X^{(u)}(\a,\bh{\k}_a^{\star}) & = X^{(s)}(\k,\bh{\a}_k^{\star}) \, , \\
	\label{eq:app_sym.stilde}
	X^{(u)}(\b,\bh{\k}_b^{\star}) & = X^{(\wt{s})} (\k,\bh{\a}_k^{\star}) \, .
	\end{align}
\end{subequations}
Given these relations, Eq.~\eqref{eq:app_sym.def} can also take the form
\begin{align}
	\label{eq:app_sym.usstilde}
	X(\k,\a) = X^{(u)}(\k,\bh{a}_k^{\star}) + X^{(s)}(\k,\bh{a}_k^{\star}) + X^{(\wt{s})}(\k,\bh{a}_k^{\star}) \, .
\end{align}

Each of these individual amplitudes have an expansion in terms of the pair angular momentum, which can be expressed in terms of any of the $u$, $s$, or $\wt{s}$ amplitudes, \eg
\begin{align}
	X^{(u)}(\k,\bh{\a}_k^{\star}) & = \sqrt{4\pi}  \, \sum_{\ell,m_{\ell}} X_{\ell m_{\ell}}^{(u)}(\k)  \,Y_{\ell m_{\ell}}^{*}(\bh{\a}_k^{\star}) \, ,\nn \\
	& = \sqrt{4\pi}  \, \sum_{\ell,m_{\ell}} X_{\ell m_{\ell}}^{(s)}(\a)  \,Y_{\ell m_{\ell}}^{*}(\bh{\k}_a^{\star}) \, .
\end{align}
In the second line, we used the relation Eq.~\eqref{eq:app_sym.s} to express the expansion in a different pair angular momentum. Since interchange of the particles in the pair affect the orientation by flipping the direction, \eg $\bh{a}_k^{\star} \to -\bh{a}_k^{\star}$ under interchange, the resulting partial wave amplitudes exhibit the symmetry condition
\begin{align}
	\label{eq:app_sym.parity}
	X_{\ell m_{\ell}}^{(\wt{s})}(\k) = (-1)^{\ell} X_{\ell m_{\ell}}^{(s)}(\k) \, .
\end{align}

For $X_{\ell m_{\ell}}^{(u)}$ representing contributions from short-distance physics, \eg the $\Kc_{3,\df}^{(u,u)}$ matrix, it is convenient to define a \emph{symmetric} function in the $k\ell m_{\ell}$ basis
\begin{align}
	\label{eq:app_sym.Xsymm}
	X_{\ell m_{\ell}}(\k) \equiv X_{\ell m_{\ell}}^{(u)}(\k) + X_{\ell m_{\ell}}^{(s)}(\k) + X_{\ell m_{\ell}}^{(\wt{s})}(\k) \, .
\end{align}
The last two terms of Eq.~\eqref{eq:app_sym.Xsymm} can be simplified by using Eq.~\eqref{eq:app_sym.parity} as
\begin{align}
	\label{eq:app_sym.Xcomb}
	X_{\ell m_{\ell}}^{(s)}(\k) + X_{\ell m_{\ell}}^{(\wt{s})}(\k)  =
	\begin{cases}
      \, \, 2X_{\ell m_{\ell}}^{(s)}(\k) & \ell \mathrm{\,\,even} \\
      \, \, 0 & \mathrm{otherwise} \, .
	\end{cases}
\end{align}
Introducing the symmetric version of $X_{\ell m_{\ell}}$ allows us too manipulate quantities which sum over all spectators in the following manner,
\begin{align}
	\label{eq:app_symm.Xo3}
	\sqrt{4\pi} \sum_{\k\in \Pc_k} \sum_{\ell,m_{\ell}} X_{\ell m_{\ell}}^{(u)}(\k) Y_{\ell m_{\ell}}^{*}(\bh{\a}_k^{\star}) & = X^{(u)}(\k,\bh{a}_k^{\star}) + X^{(s)}(\k,\bh{a}_k^{\star}) + X^{(\wt{s})}(\k,\bh{a}_k^{\star}) \, , \nn \\
	& = \sqrt{4\pi} \sum_{\ell,m_{\ell}} X_{\ell m_{\ell}}(\k) Y_{\ell m_{\ell}}^{*}(\bh{\a}_k^{\star}) \, , \nn \\
	& = \sqrt{4\pi} \sum_{\k \in \Pc_k} \sum_{\ell, m_{\ell}} \frac{1}{3}  X_{\ell m_{\ell}}(\k) Y_{\ell m_{\ell}}^{*}(\bh{\a}_k^{\star}) \, .
\end{align}
In the first line, we identified the sum over all partial wave expanded asymmetric functions with Eq.~\eqref{eq:app_sym.usstilde}, while in the second line we partial wave expand each kernel in terms of spherical harmonics of $\bh{\a}_k^{\star}$, and equate the sum of all projected functions with Eq.~\eqref{eq:app_sym.Xsymm}. The last line follows from the second, using the redundancy of the symmetric $X_{\ell m_{\ell}}$, where the factor of $1/3$ accounts for the over counting from summing over all possible spectators. Therefore, whenever the asymmetric function is contracted with the appropriate spherical harmonics and all spectators are summed over, then we may replace the asymmetric kernel with the symmetric function via the interchange
\begin{align}
	\label{eq:app_sym.Xsymm_final}
	X_{\ell m_{\ell}}^{(u)} (\k) \longrightarrow \frac{1}{3} X_{\ell m_{\ell}}(\k) \, .
\end{align}

A final identity concerning the asymmetric functions concerns with trading a $(u)$ function convoluted with an exchange operator with and $(s)$ function conjuncted with a direct channel phase space factor. The identity is given by
\begin{align}
	\label{eq:app_sym.identity2}
	\int_{p} \pi \delta & \left( (P_k - p)^2  - m^2 \right) H(\sigma_p)H(\sigma_k)\sum_{\ell,m_{\ell}} X_{\ell m_{\ell}}^{(u)}(\p)  \, 4\pi Y_{\ell m_{\ell}}^{*}(\bh{\k}_p^{\star}) Y_{\ell' m_{\ell'}}(\bh{\p}_k^{\star}) \nn \\
	& = \sum_{\ell,m_{\ell}} X_{\ell m_{\ell}}^{(s)}(\k) H(\sigma_k)  \int_{p} \pi \delta\left( (P_k - p)^2 - m^2 \right) H(\sigma_p) \, 4\pi Y_{\ell m_{\ell}}^{*}(\bh{\p}_k^{\star}) Y_{\ell' m_{\ell'}}(\bh{\p}_k^{\star}) \, ,\nn \\
	& = 2iX^{(s)}(\k) \, \wt{\rho}(\sigma_k) \, .
\end{align}
where $\wt{\rho}(\sigma_k ) = H(\sigma_k) \, \Ic(\sigma_k)$ with $\Ic$ being defined in Eq.~\eqref{eq:Ifcn}. In going from the first to second equality, the symmetry relation Eq.~\eqref{eq:app_sym.s} is used to convert the $X$ function from a spectator in $p$ to $k$. The last line is found by evaluating the integrals in the CM frame of the system of the pair associated with spectator $k$, and setting $H(\sigma_p) = 1$ since the integration of the delta function picks out the physical point.

\bibliography{bibi.bib}

\end{document}